\documentclass[a4paper,12pt]{article}

\usepackage{authblk}
\usepackage{hyperref}
\usepackage{braket}
\usepackage{amssymb}
\usepackage{amsmath}
\usepackage{amsthm}
\usepackage{amscd}
\usepackage[all,arc,knot,matrix]{xy}

\usepackage{graphicx}
\usepackage{geometry}
\newtheorem{theorem}{Theorem}[section]

\theoremstyle{definition}
\newtheorem{definition}[theorem]{Definition}

\newtheorem{construction}[theorem]{Construction}

\theoremstyle{remark}

\begin{document}

\title{Tensor Networks for Quantum Causal Histories}

\author[1]{Xiao-Kan Guo\footnote {E-mail: kankuohsiao@whu.edu.cn }
}
\affil[1]{Department of Physics, Beijing Normal University, Beijing 100875, China}

\date{\today}

\maketitle

\begin{abstract}
In this paper, we construct a tensor network representation of quantum causal histories, as a step towards directly representing states in quantum gravity via bulk tensor networks.  
Quantum causal histories are  quantum extensions of causal sets in the sense that on each  event in a causal set is assigned a Hilbert space of quantum states, and the local causal evolutions between events are modeled by completely positive and trace-preserving maps. Here we utilize the channel-state duality of completely positive and trace-preserving maps to transform the causal evolutions to bipartite entangled states. We construct the matrix product state for a single quantum causal history by projecting the obtained bipartite  states onto the physical states on the events. We also construct the two dimensional tensor network states for entangled quantum causal histories in a restricted case with compatible causal orders. The possible holographic tensor networks are explored by mapping the quantum causal histories in a way analogous to the exact holographic mapping. The constructed tensor networks for quantum causal histories are exemplified by the non-unitary local time evolution moves in a quantum system on temporally varying discretizations, and these non-unitary evolution moves are shown to be necessary for defining a bulk causal structure and a quantum black hole. Finally, we comment on the limitations of the constructed tensor networks, and discuss some directions for further studies  aiming at applications in quantum gravity.

\end{abstract}
\thispagestyle{empty}
\newpage
\pagenumbering{arabic}
\section{Introduction}
Tensor networks are efficient combinatoric descriptions of many-body quantum states in various areas ranging from condensed matter to gravity. (See \cite{BC17} for a recent introduction to tensor networks.) In particular, the entanglement properties of  many-body systems become  accessible from the network structures. This feature leads us to new understandings of the results in quantum information and computation. Meanwhile, since tensor networks arise as efficient (numerical) algorithms to extract physically relevant information from the whole many-body Hilbert space of exponentially large dimension, they are very suitable for renormalization group analyses, the striking results of which are the applications in gauge/gravity duality using the multi-scale entanglement renormalization ansatz (MERA) \cite{holography}. An interesting combination of the above two features makes it possible to study the holographic entanglement entropy and holographic error correction codes through random tensor networks \cite{HNQTWY16}. It is not difficult to see from the holographic duality that there should be a tensor network representation for the  spacetime geometry in addition to the tensor network representation of quantum states. Such tensor networks for spacetime geometries are usually studied in the anti-de Sitter/ conformal field theory (AdS/CFT) correspondence with a tacit  tensor network of quantum states in the corresponding quantum field theory (QFT). (See, for example, the new developments \cite{BPSW19} and references therein.)  

Although the holography principle appears to be a generic conviction in different models of quantum gravity, in most of these models one does not work in an explicit holographic context, even if they are indeed holographic. If one neglects the dual QFT and restricts oneself to the gravity side, then the tensor network representation of (quantum) gravity should still hold.
This raises the question of constructing tensor networks for quantum gravity without guidances from a boundary QFT. Relevant to this question is the recent work \cite{COZ18} relating the random tensor networks to the spin network structure in the  group field theory  approach to loop quantum gravity. The spin networks are quantum gravitational excitations and carry the discrete geometric data of quantum spacetime. However, after contracting the tensors living on a spin network, one obtains a tensorial group field theory on the boundary of the spin network, and the dynamical bulk quantum spacetime, i.e. spin foams, is obtained by expanding the boundary group field theory correlation functions into Feynman graphs. So such a construction of tensor networks for quantum spacetime is still a holographic one. Besides, the special relations between group field theories and spin foams are  model-dependent, and they are not applicable to other models of quantum gravity. But we can see from this construction that we need to know what the tensors represent and what we mean by the tensor contractions so as to represent a spacetime structure with tensor networks.

In this paper, we would like to study tensor network representations of general quantum spacetimes. Here by quantum spacetimes we mean that, on the one hand, the pertinent geometry is dynamically described by a quantum theory of gravity, and on the other hand, the spacetime structures such as causality and locality remains to hold in  quantum gravity. A recent impetus for studying such quantum spacetime tensor networks comes from the paper \cite{YBCHN19} where a causal tensor network is proposed for a Lorentzian spacetime. In \cite{YBCHN19}, the causal structure of a Lorentzian spacetime is encoded in a tensor network in the way of algebraic QFT. That is, the causal structures are implemented as the local unitaries between operator algebras defined on different parts of the tensor network. In particular, a global causal structure requires that the underlying graph of the tensor network is an oriented acyclic graph. This graphical structure is similar to causal sets (causets) \cite{BLMS87}. 
But   only local causal unitaries are considered in \cite{YBCHN19} and the   causal structure of  Cauchy slices in a Lorentzian spacetime is built from local unitaries, while in canonical quantum gravity one usually hopes to unitarily evolve a spacelike Cauchy surface {\it in extenso} and then describe the local evolutions by some not necessarily unitary subsystem dynamics. In tensor networks, the Trotter tensor network is exactly the case in which a global unitary evolution  can be decomposed into local unitaries via a Suzuki-Trotter decomposition. The local unitaries still need to be written in the Choi-Jamio\l kowski representation so as to be glued together to form a tensor network \cite{CHQY18}. 

Here we are interested in the case where the local evolutions of histories are not unitary, because it is not clear whether the local dynamics of quantum gravity should come from a Trotterization of a global unitary one. But we hope the Choi-Jamio\l kowski representation might still hold.
 To  this end, we shall consider in the following an old theory of quantum causal histories \cite{Mar00,EMS03}. A quantum causal history is a sequence of events (or a history) in a causet with a Hilbert space of quantum states assigned to each event. The dynamical evolutions between the quantum states on causally related events are unitary evolutions if the events at an equal time form a complete set of events as a discrete analog of spacelike Cauchy surface. However, for a single event, the causal evolutions of the states on it can be effectively described by completely positive and trace-preserving (CPTP) maps \cite{Jam72,Cho75}, since these states on a single event are in effect a subsystem in a complete set of events. The CPTP maps have the property known as the channel-state duality \cite{JLF13}, with which we can treat a CPTP map as a bipartite state in the composite system consisting of the two Hilbert spaces on the initial and final events. In this way, each CPTP map connecting a pair of causally related events can be changed to a bipartite state $\ket{\phi_e}$ living on the causal link $e$. By further specifying the state $\ket{M_v}$ in the Hilbert space over an event at the node $v$, we obtain the following formal definition of the tensor network representation for quantum causal history states (cf. Definition \ref{3311})
\begin{equation}
\ket{\mathsf{h}}=\bigotimes_e\bra{\phi_e}\bigotimes_v\ket{M_v},
\end{equation}
which has the same form as in previous works \cite{HNQTWY16,COZ18}.

In Sec. \ref{sec3}, we give an explicit construction of the tensor network state of  a single quantum causal history. The procedure is similar to the construction of matrix product states (MPS) from one-dimensional projected entangled pair states (PEPS). We also show that these tensor network states of quantum causal histories contribute a history-dependent weighting factors to the transition amplitude of that history. Unlike the consistent histories in the history approach to quantum theory \cite{Gri84}, the single-history tensor network states are themselves entangled states. This major difference is due to the fact that the causal relations between events are now translated into the entanglement between the states on events via the channel-state duality. So the extra weights can be easily obtained by contracting the tensorial coefficients without referring to the decoherence functionals.

Classically, the dynamical evolutions on a causet can be described by classical stochastic processes, such as the Markov processes in the sequential growth dynamics \cite{RS99}. Now for quantum causal histories, the dynamical evolutions of the subsystem of events are described by CPTP maps, which are  quantum Markov processes. Now that the history states are quantum, they can be quantum mechanically superposed, leading to entangled quantum histories \cite{CW16}. In Sec. \ref{sec3}, we  define a version of entangled quantum causal histories that preserves the causal order in a single history. This definition, though restrictive, avoids the complication of indefinite causal structure, so that we will be safe in working with causets. We also discuss possible ways to formulate the holographic entanglement renormalization in quantum causal histories.

Quantum causal histories underlie the causal evolutions in many models of quantum gravity. However, in most of the existing examples, such as  spin networks \cite{MS97},   strings \cite{MS98} and spin foams \cite{LO03}, the causal evolutions are unitary, even if they are local. Although there are strong reasons to study a model of quantum gravity  by local unitaries (see for example \cite{AM17}), the possible discretization or graph changing dynamics will lead to non-unitary evolutions. In Sec. \ref{sec4}, an example of non-unitary local time evolutions for a quantum system on temporally varying discretizations \cite{Hoh14} is discussed in comparison with quantum causal histories. An explicit MPS is constructed for a history of local evolutions that includes the non-unitary ones.  These non-unitary local evolution moves are important for describing a local causal structure and hence a black hole region in the related tensor network, thereby making the constructed tensor networks physically more relevant.

In Sec. \ref{sec5}, we  comment on the potential weakness in our construction in view of the lack of examples in most existing models. We also point out several  directions of further studies.

In Appendix \ref{AppA}, we show that the tensor network state of a single quantum causal history is indeed a superstate in spacetime quantum mechanics \cite{CJQW18}. In Appendix \ref{AppB}, we discuss some categorical aspects of quantum causal histories and their tensor netwrok representations.
\section{Quantum causal histories revisited}
In this section, we first recollect the essentials of the theory of quantum causal histories. The basic references are \cite{Mar00,EMS03}. Then we discuss two new assumptions on quantum causal histories: The first is imposing the internal causal condition for the evolutions overlapping at a single event, which ensures the background independence of the sum over histories (Sec. \ref{2.2}); the second is  choosing the reference states such that the channel-state duality can be applied in quantum causal histories (Sec. \ref{2.3}).
\subsection{The old story: quantum causal histories}
The phrase ``quantum causal history" reveals itself the content of this theory: by ``quantum" it means a quantum  theory for gravitational evolutions with the existence of well-behaved Hilbert spaces and evolutions assumed; by ``causal" it means the causet approach to quantum gravity  where the causal evolutions between  events are collected into a partially ordered set (poset); finally, by ``history" it means the history formulation of quantum theory  that can describe a sequence of events. Altogether,   quantum causal histories are causets with Hilbert spaces on the events and there are evolution operators mapping between these Hilbert spaces.

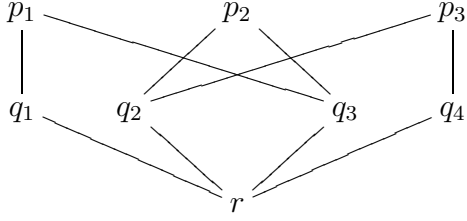
\begin{figure}[t]
\[
\xymatrix{
p_{1} \ar@{-}[d] \ar@{-}[drrr] & &
p_{2} \ar@{-}[dl] \ar@{-}[dr] & &
p_{3} \ar@{-}[dlll] \ar@{-}[d] \\
q_{1} \ar@{-}[drr] &
q_{2} \ar@{-}[dr] &&
q_{3} \ar@{-}[dl] &
q_{4} \ar@{-}[dll] \\
& & r & & }
\]
\caption{An example of causet. The direction of causal evolution is from bottom to top.}\label{f1}
\end{figure}

More explicitly, consider a finite set $\mathcal{C}\equiv\{p,q,r,...\}$ of (spacetime) events  labelled by points $p,q,r,...$. For two events $p$ and $q$, we denote by $p\leqslant q$ the condition that $p$ causally precedes $q$. Then the set  $\mathcal{C}$ is a poset. $\mathcal{C}$ becomes a causet if the causal relation $\leqslant$ is reflexive ($p\leqslant p$), antisymmetric (if $p\leqslant q$ and $q\leqslant p$, then $p=q$), and transitive (if $p\leqslant q$ and $q\leqslant r$, then $p\leqslant r$). The antisymmetry condition precludes the closed timelike curves.
A sequence of events in $\mathcal{C}$ related by the causal relation $\leqslant$ form a causal history (or chain) $\mathsf{h}(p)$.
Let us denote by $P(p)$ the causal past of the event $p$, i.e. all events $r\in\mathcal{C}$ such that $r\leqslant p$. Likewise, denote by $F(p)$ the causal future of $p$ consisting of all $q\in\mathcal{C}$ such that $p\leqslant q$. The causal past (or future)  of $p$ consists of many (past/future oriented) causal histories, e.g. $P(p)=\bigcup_i\mathsf{h}_i(p)$, but we should keep in mind that these causal histories may intersect each other. For example, in Figure \ref{f1} the causal past $P(p_1)$ comprises the causal histories $\{r\leqslant q_1\leqslant p_1\}$ and $\{r\leqslant q_3\leqslant p_1\}$, but $\{r\leqslant q_3\leqslant p_1\}$ intersects $\{r\leqslant q_3\leqslant p_2\}$.

Not all events in $\mathcal{C}$ can have causal relations. Those events $a,b,c,...\in\mathcal{C}$ that cannot be causally related to each other are said to be spacelike separated, and they  form an acausal set ${A}\subset\mathcal{C}$. 
Using acausal sets, we can in some sense complimentarily characterize the causal events. To see this, we call an acausal set $A$  a complete past of an event $p$ if  $A\cap\mathsf{h}_i(p)\neq\emptyset$ for each causal history $\mathsf{h}_i(p)$, and likewise for a complete future. For two acausal sets $A$ and $B$, define a relation $A\preceq B$ to be the condition that $A$ is the complete past of $B$ and $B$ is the complete future of $A$. We call such a pair $A,B$ with $A\preceq B$ a complete pair, which is the discrete analog of a pair of successive Cauchy surfaces.
 It is easy to check that the relation $\preceq$ is reflexive, anti-symmetric and transitive. So we denote the poset of acausal sets by $\mathcal{A}$. An example is given in Figure \ref{f1} where the acausal sets $A=\{q_1,q_2,q_3,q_4\}$ and $B=\{p_1,p_2,p_3\}$ form a complete pair.

Next, let us assign a finite-dimensional Hilbert space $H(q)$ to each event $q\in\mathcal{C}$. For an acausal set $A=\{q_1,q_2,...,q_n\}$, the Hilbert space of $A$ is the tensor product of event Hilbert spaces, $H(A)=\bigotimes_iH(q_i)$, since the events in $A$ are spacelike separated with respect to each other. Similarly, if two acausal sets $A$ and $B$ are spacelike separated, we have the composite Hilbert space $H(A)\otimes H(B)$.

 For two acausal sets $A,B\in\mathcal{A}$ that are partially ordered as $A\preceq B$,  we assign an evolution operator $E_{AB}:H(A)\rightarrow H(B)$ to their Hilbert spaces. If  furthermore  $\dim H(A)=\dim H(B)$, then $E_{AB}$ is a unitary operator. Here the unitary evolution  operator need not to be generated by a physical (gravitational) Hamiltonian because of the discreteness of the time steps, although it is quite possible to have relations to the Hamiltonian in continuous time.
Now the properties of the relation $\preceq$ can be expressed through these evolution operators, e.g. the reflexive property is $E_{AA}={\bf1}_A$ and the transitive property is $E_{AB}E_{BC}=E_{AC}$. 

 Notice that the unitary evolutions on acausal sets hide the causal relations between individual events in a causet. 
To amend this, consider the $C^*$-algebra $\mathcal{U}(p)$ of linear operators acting on $H(p)$, and similarly for an acausal set $A$ we have  $\mathcal{U}(A)=\bigotimes_i\mathcal{U}(p_i)$. Here the involutions in a $C^*$-algebra are defined by  the inner products in the assigned Hilbert space. Then a $C^*$-algebra $\mathcal{U}(B)$  can be unitarily changed to $\mathcal{U}(A)=E_{AB}^\dag\mathcal{U}(B)E_{AB}$ through a $^*$-isomorphism, but the $E_{AB}$'s cannot be used to evolve an algebra $\mathcal{U}(p)$ on a single event $p$. To define consistent maps between the $C^*$-algebras  on events, we use the CPTP maps  between $C^*$-algebras. Intuitively, an evolution map $E_{pq}:H(p)\rightarrow H(q)$ from  $H(p)$ on an event $p$ to  $H(q)$ with $p\in A,q\in B$ and $A\preceq B$ is the evolution of the subsystem in a complete acausal set.
The CPTP maps are then natural descriptions of such subsystem evolutions in open quantum systems. 
Here the map $E_{pq}$ should be a map of density matrices which are states on matrix algebras. The use of matrix algebras is due to the fact that the  $C^*$-algebras acting on the assigned finite-dimensional Hilbert space are type I von Neumann factors.

To relate the CPTP maps on $C^*$-algebras to the CPTP maps on states, let us consider the following:
 On a type I  factor as a matrix algebra there is a unique trace $\tau=\text{tr}$, i.e. the normalized matrix trace. 
The trace $\tau$ defines an inner product on the $C^*$-algebra $\mathcal{U}(p)$, e.g. $\tau(a^*b)$ for $a,b\in\mathcal{U}(p)$, and hence the $C^*$-algebras $\mathcal{U}(p)$ are also Hilbert spaces. We denote the algebraic involution in $C^*$-algebras by $^*$ and the adjoint of map in the $C^*$-algebra as Hilbert spaces by $^\dag$. Then the evolution map $E_{pq}$ is indeed the adjoint of the map $\phi(pq)$ on operator algebras, $E_{pq}=\phi^\dag_{pq}$ where $\phi_{pq}:\mathcal{U}(q)\rightarrow\mathcal{U}(p)$.
With these notations, we have the following local definition according to \cite{EMS03}:
\begin{definition} \label{d21}
A quantum causal history $\mathsf{h}$ is a sequence of events in a causet $\mathcal{C}$, where to each event $p\in\mathcal{C}$ is assigned a type I von Neumann factor $\mathcal{U}(p)$ and to every pair of causally related  events $p\leqslant q$  is assigned a CPTP map $\phi_{pq}:\mathcal{U}(q)\rightarrow\mathcal{U}(p)$, satisfying the following conditions:
\begin{enumerate}
\item {\it Extension}. For any $q\in\mathcal{C}$, there exists a homomorphism $\phi_{Pq}:\mathcal{U}(q)\rightarrow\mathcal{U}(P)$, where $P$ is the complete past of $q$, such that for each $p\in P$ it reduces to a CPTP map $\phi_{pq}$. Likewise, for a complete future $R$ of $q$, there exists a homomorphism $\phi_{qR}^\dag:\mathcal{U}(q)\rightarrow\mathcal{U}(R)$ such that for each $r\in R$, $\phi_{qR}$ reduces to the CPTP map $\phi_{qr}$.
\item {\it Locality}. If $p,q\in\mathcal{C}$ are spacelike separated, then for the complete past $A$ of $p,q$ the images of $\phi_{Ap}$ and $\phi_{Aq}$ commute. Likewise, for the complete future $B$ of $p,q$, the images of $\phi_{pB}^\dag$ and $\phi_{qB}^\dag$ commute.
\item{\it Transitivity}. If $C$ is a complete future of $p$ and a complete past of $q$, then $\phi_{pq}=\phi_{pC}\phi_{Cq}$.
\end{enumerate}
\end{definition}
The assignment of von Neumann algebras to spacetime events does not allow us to formulate  a complete QFT on quantum causal histories. We can only have parts of the properties of an algebraic QFT.
To  this end,  let us define for a subset $X\subset\mathcal{C}$ its causal complement  $X'$ as the set of events that are spacelike to all of $X$. Then the causal completion of $X$ is $X^{\prime\prime}$, and $X$ is causally complete if $X=X^{\prime\prime}$. Given an acausal set $A$, we have $\mathcal{U}(A^{\prime\prime})=\mathcal{U}(A)=\bigotimes_{a\in A}\mathcal{U}(a)$. Then for another acausal set $B$ such that $A$ is the complete past of $B$, we have the isotony of algebras $\mathcal{U}(A^{\prime\prime})\subset\mathcal{U}(B^{\prime\prime})$. We thus obtain the two most important properties of algebraic QFTs that are relevant to tensor network constructions, namely the (Einstein) locality and isotony of algebras, cf. \cite{YBCHN19}.

\subsection{Interlude: background independence}\label{2.2}
The quantum causal histories defined above constitute a discrete, causal, quantum and finite model of quantum spacetime (or QFT on quantum spacetime). However, since the causal structure is fixed, such a model is not background independent, which is important if we use this model to study quantum gravity. A direct thought is to consider unfixed histories and sum over these histories as in \cite{Mar00}. Formally, given a complete pair $A\preceq B$, we denote by $\gamma(A,B)$ the graph underlying all the causal histories between $A$ and $B$. Now if $\mathcal{A}$ is not a fixed set, then on each distinct graph $\gamma$ in between we can assign an evolution operator $E^\gamma_{AB}$. The transition amplitude from a state $\ket{\psi_A}\in H(A)$ to $\ket{\psi_B}\in H(B)$ on this graph $\gamma$ is obviously $\braket{\psi_B|E^\gamma_{AB}|\psi_A}$, and the total transition amplitude is obtained by a sum over histories 
\begin{equation}\label{1111}
\mathsf{A}(A,B)=\sum_\gamma\braket{\psi_B|E^\gamma_{AB}|\psi_A}.
\end{equation}

It is unclear whether the histories summed over are still causal, because the causal structures inside these histories are invisible in the final transition amplitude even if one starts with a fixed quantum causal history. In \cite{Wal13} the causal evolutions in a sum over histories is prescribed by borrowing  the ideas of prescribing the race condition and deadlock  in parallel programming. The main idea (of race conditions), when stated in the current case of quantum causal histories, is that at an event there might be more than one possible future-oriented histories, which poses the question of which causal evolution happens ``first". The prescription is then  to simply restrict ourself to non-overlapping histories by taking as granted that the causal evolutions that overlap on the same event have been properly ordered.  

\begin{figure}[t]
\[
\xymatrix{
q_{1} \ar@{-}[dr] &&q_{2} \ar@{-}[dl]&~\ar @{} [dl] |{\Longrightarrow}\\
 & p & }
\xymatrix{
q_{1} \ar@{-}[d] &
q_{2} \ar@{-}[d] \\
p_1\ar @{} [r] |{\curlyeqprec} & p_2 }
\]
\caption{The assumption of internal causality.}\label{f2}
\end{figure}
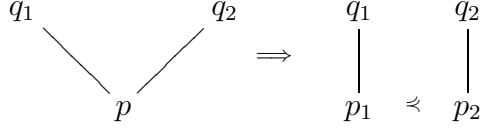

Let $\mathcal{Q}_\mathcal{C}$ be a set of quantum causal histories based on a causet $\mathcal{C}$, then for two or more quantum causal histories $\mathsf{h}_i,i=1,2,...$ coinciding at an event $p\in\mathcal{C}$, we introduce a new causal ordering   $\mathsf{h}_i(p)\curlyeqprec\mathsf{h}_j(p)$ at $p$, or equivalently we change the event $p$ into many copies ordered as $p_1\curlyeqprec p_2$ etc.. With this new ordering at an overlapping event, these histories become disjoint. For example, consider two quantum causal histories $\mathsf{h}_1,\mathsf{h}_2$ coinciding only at  $p$, we change them to ${\mathsf{h}'_1},\mathsf{h}'_2$ with $p$ replaced  respectively by $p_1$ and $p_2$ satisfying $p_1\curlyeqprec p_2$. We have then
\[
\mathsf{h}_1\cap\mathsf{h}_2=p,~\mathcal{Q}_\mathcal{C}=\mathsf{h}_1\cup\mathsf{h}_2\quad\Rightarrow\quad\mathsf{h}'_1\cap\mathsf{h}'_2=\emptyset,~\mathcal{Q}_\mathcal{C}=\mathsf{h}'_1\sqcup\mathsf{h}'_2/\{p_1=p_2\}.
\]
See also Figure \ref{f2}.

This additional ordering is internal to the event $p$ and does not affect the structure of quantum causal histories in $\mathcal{Q}_\mathcal{C}$, and in a sense is hidden in the reflexivity condition $p\leqslant p$, which   replaces  the sometimes assumed irreflexivity in the literature. We call this assumed causal relation as {\it internal causality}.
Importantly, the quantum causal histories can still have overlaps and   even can be entangled. This internal causality makes the quantum causal histories into an enriched category, cf. Appendix \ref{AppB}.

\subsection{A new story: channel-state duality}\label{2.3}
The use of  CPTP maps as local evolution maps in quantum causal histories allow us to utilize tools from quantum information theory, since the CPTP maps are actually a class of quantum channels \cite{JP18}.\footnote{Usually  quantum channel is a synonym for CPTP map. But there are non-CP maps which  can be quantum channels with initial system-environment correlations.}
 This prospect has been explored in \cite{LT07} where possible extensions to non-CP maps are also discussed. Notice that  non-CP maps or affine maps typically arise when there are initial system-environment correlations. By the assumed internal causality, the quantum causal histories  overlapping at an event should be causally ordered in a way  that these histories are effectively disjoint. 
So, it is reasonable to consider only the cases without initial correlations, and treat  the local causal evolutions in quantum causal histories as CPTP maps.

Here we exploit the channel-state duality of CPTP maps \cite{JLF13}. Let $H_a$ be a finite-dimensional Hilbert space, and $B(H_a)$ be the space of bounded linear operators on $H_a$. $B(H_a)$ is also a Hilbert space with the Hilbert-Schmidt inner product $\braket{a_1|a_2}=\text{tr}(a_1^\dag a_2)$. Let $H_b$ be another finite-dimensional Hilbert space, then a CPTP map $\varphi:B(H_a)\rightarrow B(H_b)$ corresponds to a bipartite states on $H_a\otimes H_b$ by the following duality, also known as the Choi-Jamio\l kowski isomorphism \cite{Cho75,Jam72} (for unitaries), 
\begin{equation}\label{CSD}
\varphi\rightarrow\rho_\varphi={\bf1}\otimes\varphi(\ket{\psi}\bra{\psi})=\sum_{ij}e_{ij}\otimes\varphi(e_{ij}),
\end{equation}
where $\ket{\psi}=\sum_i\ket{i}\otimes\ket{i}$ is a maximally entangled state in $H_a\otimes H_a$ and $e_{ij}=\ket{i}\bra{j}$ for an orthonormal basis $\{\ket{i}\}$ of $H_a$.  The dual states obtained by the map \eqref{CSD} belong to a Hilbert subspace of the Hilbert space $H_a\otimes H_b$, because the CPTP map $\varphi$ preserves the Hilbert space structure. 

In the current case, the CPTP map $\phi_{pq}^\dag$ mapping the density matrix in $H(p)$ to that in $H(q)$ is dual to a bipartite state 
\begin{equation}\label{3}
\phi_{pq}^\dag\rightarrow\rho_{pq}={\bf1}\otimes\phi_{pq}^\dag(\ket{\psi}\bra{\psi}).
\end{equation}
For later use, we also consider the images of $\phi_{pq}^\dag$ in terms of state vectors. By noticing that the map $\varphi:(e_{ij})_a\mapsto\varphi(e_{ij})=(e_{ij})_b$ in \eqref{CSD} changes the basis vectors of $H_a$ to those of $H_b$, 
 we can write the corresponding transformations of state vectors obtained by channel-state duality as
\begin{align}
\ket{\psi_{pp}}=&\sum_{p}\ket{p}\otimes\ket{p}\mapsto\ket{\psi_{pq}}=\sum_{pq}\ket{p}\otimes\ket{q}\equiv\sum_{p}\ket{p}\otimes\check{\phi}^\dag_{pq}(\ket{p})\label{dup}\\
\bra{\psi_{pp}}=&\sum_{p}\bra{p}\otimes\bra{p}\mapsto\bra{\psi_{pq}}=\sum_{pq}\bra{p}\otimes\bra{q}\equiv\sum_{p}\bra{p}\otimes\hat{\phi}^\dag_{pq}(\bra{p}) \label{du4}
\end{align}
for $\ket{p}\in H(p),\ket{q}\in H(q)$ and $\bra{p}\in H^*(p),\bra{q}\in H^*(q)$.

Now we face the problems of choosing a pair of reference states $\ket{\psi}$ on each causal link and choosing the basis $\{\ket{i}\}$ on each event. For the reference  state, we can still choose a maximally entangled state on the condition that the (reference) basis of $H(p)$ has been chosen. This choice is consistent with gluing of polyhedra in loop quantum gravity. Although this point is not manifest in most of the approaches to loop quantum gravity, it is manifest in the so-called bosonic representation of squeezed vacua \cite{BBY18} that the gluing of two polyhedra requires the entanglement between  two intertwiners to be maximal.\footnote{We also note that the bosonic representation of squeezed vacua is closely related to  the spinor representation of loop quantum gravity \cite{LT12} where the Hilbert space of quantized spinors are the Bargmann space of holomorphic square integrable functions over
$\mathbb{C}$. An interesting thing pointed out in \cite{JLF13} is that the inverse channel-state duality, namely constructing a channel $\varphi_\sigma(\cdot)=\text{tr}_a[(\cdot\otimes{\bf1})\sigma]$ from a given state $\sigma$, is also formally analogous to the Segal-Bargmann transform $B(\cdot)=\int dx(\cdot) \ket{z,x}$ from  coherent states $ \ket{z,x}\in L^2(\mathbb{R}^2)$ to holomorphic functions on the Bargmann space. We therefore expect an exact formal correspondence between the spinor representation of loop quantum gravity and state-channel duality, which is worthy of further investigations.}
The CPTP maps represent the dynamical evolutions of events that can possibly change the maximal entanglement. In loop quantum gravity terms, these evolution maps can be understood as propagating curvature excitations along the edges that originally maximally entangle the polyhedra in the squeezed vacua. 

As for choosing the basis of reference,  the problem does not lie in the existence of  an orthonormal basis for a finite-dimensional Hilbert space, but in the fact that the state in \eqref{CSD} becomes basis-dependent when we change the basis using local operations \cite{JLF13}. Here we choose the  basis of the Hilbert space $H(p)$ at an event $p\in A$ according to other events in the acausal set $A$. In this sense, the fixation of basis by choosing other events becomes similar to the problem of choosing a suitable quantum reference frame in a composite system, but with a notable difference:
 In canonical quantum gravity, it is required that a choice of quantum reference frame should give the physical unitary Schr\"odinger dynamics of quantum gravity, whereas here we work with an event in an acausal set (i.e. a point in a Cauchy surface) and its open system dynamics rather than unitary dynamics. To avoid further complications, we simply assume that such a choice of orthonormal basis for the Hilbert space assigned to each event is not only unique but also consistent with the evolutions of acausal sets.

\section{Tensor networks for quantum causal histories}\label{sec3}
We construct in this section the tensor network representation of quantum causal histories. For a single quantum causal history, its tensor network representation of MPS form is first constructed in Sec. \ref{3.1}. 
We then discuss some properties of these MPS on a set of histories in Sec. \ref{3.x}. In Sec. \ref{3.2}, we define tensor networks for entangled quantum causal histories, but with their causal structures kept intact. And finally we discuss in Sec. \ref{3.3} the possible ways to realize holography in quantum causal histories with the help of tensor network representations.
\subsection{Matrix product state of a single history}\label{3.1}
To construct a tensor network representation of a single quantum causal history, let us take a step back and look at  the history states in the theory of consistent histories \cite{Gri84,Ish94}. For a quantum system whose dynamics is given by the unitary operators $U$, consider a finite sequence of projection operators $\alpha_{t_1},\alpha_{t_2},...,\alpha_{t_n}$ corresponding to quantum logical propositions made at different times $t_1,t_2,...,t_n$. Then the history states of this quantum system are states in the Hilbert space $\mathcal{H}=\bigotimes_iH(t_i)$ where $H(t_i)$ is the Hilbert space of the system at time $t_i$. In particular, a history $\mathsf{h}$ in $\mathcal{H}$ is projected out by a class operator 
\begin{equation}
C_{\mathsf{h}}=U^\dag (t_n)\alpha_{t_n}U(t_n)... U^\dag (t_1)\alpha_{t_1}U(t_1).
\end{equation}
For two histories $\mathsf{h}_1$ and $\mathsf{h}_2$, their decoherence functional is $D(\mathsf{h}_1,\mathsf{h}_2)=\text{tr}(C_{\mathsf{h}_1}\rho_0C^\dag_{\mathsf{h}_2})$ where $\rho_0$ is the initial density matrix of the system. The decoherence functional gives the probability distribution $p(\mathsf{h})=D(\mathsf{h},\mathsf{h})$ of a history $\mathsf{h}$ if the consistency condition, $D(\mathsf{h}_1,\mathsf{h}_2)=0$ for $\mathsf{h}_1\neq\mathsf{h}_2$, is satisfied.

Now for quantum causal histories, we see two major differences when compared to consistent histories: (i) the dynamics of events in a single quantum causal history are not unitary but described by CPTP maps, and consequently the Hilbert spaces on different events in a quantum causal history might be very different; (ii) in quantum causal histories the projection operators are not explicitly included. However, these two differences are closely relevant to the tensor network constructions: on the one hand, the CPTP maps can be transformed via channel-state duality to bipartite states on the ``bond" corresponds to contracting the tensor indices; on the other hand, the projection operators tell us that we need extra ``physical" indices to represent the events themselves in addition to bond indices.

Let $\mathsf{h}$ be a quantum causal history consisting of $n$ events $p_1,p_2,...,p_n$. On each $p_i, i=1,2,...,n$, we have a Hilbert space $H(p_i)$ and a type I factor $\mathcal{U}(p_i)$, and the states $\ket{\psi_i}$ in $H(p_i)$  are related by local causal evolutions or CPTP maps. In terms of density matrices, we write the causal evolutions at the level of states as a one-dimensional directed chain or graph,
\begin{equation}\label{66}
\rho_1\xrightarrow{\phi^\dag_{12}}\rho_2\xrightarrow{\phi^\dag_{23}}...\xrightarrow{\phi^\dag_{(n-1)n}}\rho_n.
\end{equation}
Each $\phi^\dag_{ii+1}$ can be changed into a bipartite entangled state as in \eqref{CSD}, which in effect puts an entangled pair on the bond of the graph. Since the graph is directed according to the causal relations in $\mathsf{h}$, the evolved state $\phi^\dag(\ket{\psi_i})$ should be put at the target node $(i+1)$ of the bound $(i,i+1)$, and the other state of the entangled pair at the source node $(i)$ of the bond. For example, consider the following part of the graph
\begin{equation}
\rho_1\xrightarrow{\phi^\dag_{12}}\rho_2\xrightarrow{\phi^\dag_{23}}\rho_3\xrightarrow{\phi^\dag_{34}}...
\end{equation}
then the effects of the maps $\phi^\dag_{12},\phi^\dag_{23}$ result in the following bipartite states
\begin{equation}
\ket{\psi_{12}}=\sum_i\ket{\psi_{1,i}}\otimes\check{\phi}^\dag_{12}(\ket{\psi_{1,i}}), \quad \ket{\psi_{23}}=\sum_i\ket{\psi_{2,i}}\otimes\check{\phi}^\dag_{23}(\ket{\psi_{2,i}}).
\end{equation}
At the event $p_2$ we insert a projection operator
\begin{equation}\label{99}
\beta_2=\sum_{ijk}M^{(2)}_{12,ijk}\ket{\psi_k^{(2)}}\hat{\phi}^\dag_{12}(\bra{\psi_{1,i}})\bra{\psi_{2,j}},
\end{equation}
where the $M$'s are some tensorial coefficients. The states $\ket{\psi^{(2)}}\in H(p_2)$ are those states that characterize $\mathsf{h}$ at the event  $p_2$ and also specifies the way of gluing two CPTP maps linked to $p_2$. Note that $\{\ket{\psi^{(2)}_k}\}$ is again a basis of $H(p_2)$ such that $\ket{\psi^{(2)}}=\sum_kM^{(2)}_k\ket{\psi^{(2)}_k}$.

We emphasize that the projection operator $\beta_2$ is used to glue two causal evolutions, since the evolved state  $\check{\phi}^\dag_{12}(\ket{\psi_{1}})$ may stop at $p_2$ and the subsequent evolution is about another state $\ket{\psi_{2}}\in H(p_2)$.  This is because the channel-state duality applies only on the condition that the maximally entangled reference states are chosen. In this case, the (unnormalized) reference state is $\sum_{ij}\ket{\psi_{1,i}}\ket{\psi_{2,j}}$.
The tensorial coefficients $M_{12,ijk}$'s thus encode the transition from $\check{\phi}^\dag_{12}(\ket{\psi_{1,i}})$  to $\ket{\psi_{2,j}}$, and in the state $\ket{\psi^{(2)}}$ is encoded the pertinent property of the event $p_2$, e.g. the eigenstate of the operators from $\mathcal{U}(p_2)$.
In the simple case where $\check{\phi}^\dag_{12}(\ket{\psi_{1,i}})=\ket{\psi_{2,i}}$, we can take $\ket{\psi_i^{(2)}}=\ket{\psi_{2,i}}$ and  $M={\bf1}$ as an identity matrix. 

\begin{figure}[t]
\[
\xymatrix{
\dots\ar@{-}[r] & *+[F]{M^{(i-1)}} \ar@{-}[d]\ar@{-}[r]  &  *+[F]{M^{(i)}}\ar@{-}[d]\ar@{-}[r] &  *+[F]{M^{(i+1)}}\ar@{-}[d]\ar@{-}[r]&\dots\\ 
\dots& \ket{\psi^{(i-1)}} & \ket{\psi^{(i)}} & \ket{\psi^{(i+1)}} &\dots
 }
\]
\caption{Illustration of the MPS of a single quantum causal history. }\label{fMPS}
\end{figure}
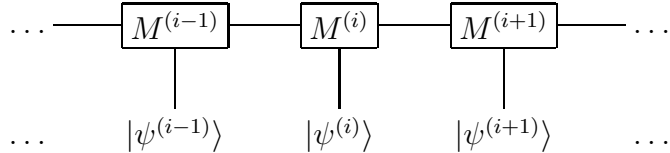

For the total graph \eqref{66}, we insert the projection operators $\beta_i,i=2,3,...,n-1$ at the corresponding events, and obtain a history state, schematically as in Figure \ref{fMPS},
\begin{align}
\ket{\mathsf{h}}=&\beta_{2}\otimes...\otimes\beta_{(n-1)}\ket{\rho_1\xrightarrow{\phi^\dag_{12}}\rho_2\xrightarrow{\phi^\dag_{23}}...\xrightarrow{\phi^\dag_{(n-1)n}}\rho_n}=\nonumber\\
=&\sum_{i_1i_2...i_n,k_2k_3...k_{n}}M^{(2)}_{12,i_1i_2k_2}M^{(3)}_{23,i_2i_3k_3}...M^{(n-1)}_{(n-1)n,i_{n-1}i_nk_n}\ket{\psi^{(1)}_{i_1}}\ket{\psi^{(2)}_{i_2}}...\ket{\psi^{(n)}_{i_n}}.\label{10}
\end{align}
This state \eqref{10} takes the form of a MPS state, and when compared to the consistent history states in $\mathcal{H}=\bigotimes_iH(t_i)$, it is obviously entangled.  
This state is also a superstate in temporally extended, or spacetime, quantum mechanics, cf. Appendix \ref{AppA}.

The above construction of the tensor network state \eqref{10}  follows the construction of MPS from one-dimensional PEPS \cite{BC17}. We can in fact  give a direct definition by relating the Hilbert spaces to the tensor indices in those $M$'s. Let the Hilbert spaces on events be $d$-dimensional. Consider the quantum causal history $\mathsf{h}$ with an underlying directed graph structure as above, then on each node $v$ of the graph there is a tripartite node state
\begin{equation}\label{V}
\ket{M_v}=\sum_{ijk}V_{ijk}^{(v)}\ket{\psi_i}\ket{\psi_j}\ket{\psi_k^{(v)}}
\end{equation}
where $\ket{\psi_i}$ and $\ket{\psi_j}$ are on the two bonds linked to the node $v$, and $\ket{\psi_k^{v}}$ is not related to the causal evolutions but resides on the node $v$. Here each index of the tensorial coefficients $V_{ijk}^{(v)}$ stands for a Hilbert space on the event $p_v$ of dimension $d$. On each bond of the graph, there is a bipartite bond state obtained from a CPTP map $\phi$ via channel-state duality. This state on the bond $e$ can be written in the dual space as in \eqref{du4},
\begin{equation}\label{E}
\bra{\phi_e}=\sum_{i}\bra{\psi_i}\hat{\phi}^\dag(\bra{\psi_i})=\sum_{ij}E_{ij}\bra{\psi_i}\bra{\psi_j}
\end{equation}
where the tensorial coefficients $E_{ij}$ encodes the transition from $\hat{\phi}^\dag(\bra{\psi_i})$ to $\bra{\psi_j}$. A tensor network state such as \eqref{10} is then obtained by contracting the tensors $V$ and $E$ (resulting in the tensorial coefficients $M$).
We can thus formulate the formal definition of the tensor  network representation of a quantum causal history:
\begin{construction}\label{3311}
Let  $\mathsf{h}$ be a quantum causal history of events $\{p_v\}$ each of which is a subsystem of the complete acausal set $A_v$.
The tensor network representation of  $\mathsf{h}$ is 
\begin{equation}
\ket{\mathsf{h}}=\bigotimes_e\bra{\phi_e}\bigotimes_v\ket{M_v}
\end{equation}
where the states $\ket{M_v}$ and $\bra{\phi_e}$ are respectively given by \eqref{V} and \eqref{E}. The bonds $e$  and nodes $v$ are in the graph underlying the history $\mathsf{h}$.
\end{construction}
If each event in $\mathsf{h}$ is a complete acausal set, then the tensor network representation can still be constructed as above, since the Choi-Jamio\l kowski isomorphism holds for a unitary transform.  Of course, one can also construct tensor networks with local unitaries along the lines of \cite{YBCHN19}. 

\subsection{Many histories}\label{3.x}
We now consider the question of summing over histories in light of the constructed tensor network representation of a single quantum causal history. In the simple expression \eqref{1111}, the amplitude for a single history  can be written  as
\begin{equation}\label{14}
\braket{\psi_n|E_{\mathsf{h}}|\psi_1}=\braket{\psi_n|\check{\phi}^\dag_{n-1}...\check{\phi}^\dag_1|\psi_1}
\end{equation}
where we have made the abbreviation the $\bf{1}\otimes\check\phi^\dag\equiv\check\phi^\dag$.
Here the composition of the channel-state duality works as
$
\phi^\dag_{n-1}...\phi^\dag_1\rightarrow\sum_{ij}e_{ij}\otimes\phi^\dag_{n-1}...\phi^\dag_1(e_{ij})
$
with each state on an event coincides with that of the maximally entangled reference state. 
This \eqref{14} is a transition amplitude  from $\ket{\psi_1}$ to $\ket{\psi_n}$, while in the tensor network  state \eqref{10} we  include the additional states $\ket{\psi^{(p)}}$ residing at the events $p$'s, so we also need to consider  some $n$-point functions of tensor network states $\ket{\mathsf{h}}$. As in the  class operators of consistent histories, we consider the $n$-point functions of the projection operators $\eta_i=\ket{\psi^{(i)}}\bra{\psi^{(i)}}, i=1,2,...,n$, to wit
\begin{equation}
\braket{\mathsf{h}|\eta_1\eta_2...\eta_{n}|\mathsf{h}}=\bigl(M^{(2)}_{12}M^{(3)}_{23}...M^{(n-1)}_{(n-1)n}\bigr)^\dag M^{(2)}_{12}M^{(3)}_{23}...M^{(n-1)}_{(n-1)n},
\end{equation}
where the right hand side is totally contracted.
This $n$-point function $\braket{\mathsf{h}|\eta_1\eta_2...\eta_{n}|\mathsf{h}}$ gives an extra weight to the transition amplitude of the quantum causal history $\mathsf{h}$. Indeed, we note that the plain transition amplitude \eqref{14} only uses the maximally entangled reference states. For states other than the reference states, we consider their transitions to the reference ones with additional tensorial coefficients,
\begin{equation}
\braket{\psi_n|M^{(n-1)}_{(n-1)n}\check{\phi}^\dag_{n-1}...M_{12}^{(2)}\check\phi^\dag_2\check\phi^\dag_1|\psi_1}=M^{(2)}_{12}M^{(3)}_{23}...M^{(n-1)}_{(n-1)n}\braket{\psi_n|\check\phi^\dag_{n-1}...\check\phi^\dag_1|\psi_1},
\end{equation}
where the tensors $M$'s have been contracted. 

We can therefore perform the weighted sum over quantum causal histories to obtain the amplitude
\begin{equation}\label{16}
\mathsf{A}(\psi_1,\psi_n)=\sum_\gamma\sqrt{\braket{\mathsf{h}_\gamma|\eta_1\eta_2...\eta_{n}|\mathsf{h}_\gamma}}\braket{\psi_n|E_{\mathsf{h}_\gamma}|\psi_1}
\end{equation}
where the internal causality condition is assumed to hold. Notice that the local operators $\eta_i$ will not pick out a special frame or direction, since the states $\ket{\psi^{(i)}}$ just encode the transition of $\mathsf{h}_\gamma$ at the events $p_i$ and the histories $\gamma$ are not fixed  in the sum-over-histories approach. Therefore, the weighted sum over histories \eqref{16} is background independent.

If we replace the projection operators $\eta_i$ by some local operator $O_i$, we get the correlation function of these operators $O_i$,
\begin{align}
\braket{\mathsf{h}|\eta_1...O_j...O_k...\eta_{n}|\mathsf{h}}=&\braket{\psi^{(j)}|O_j|\psi^{(j)}}\braket{\psi^{(k)}|O_k|\psi^{(k)}}\cdot\nonumber\\
&\cdot\bigl(M^{(2)}_{12}M^{(3)}_{23}...M^{(n-1)}_{(n-1)n}\bigr)^\dag M^{(2)}_{12}M^{(3)}_{23}...M^{(n-1)}_{(n-1)n}.
\end{align}
This correlation function now depends on the residing states $\ket{\psi^{(j)}}$ and $\ket{\psi^{(k)}}$ which are causally related by the  MPS structure of $\ket{\mathsf{h}}$. In fact, since the $\mathsf{h}$ is by definition a causal history, the operators inserted on its events are causally correlated. But unlike the usual homogeneous  MPS, the correlation strength now depends on those  states residing at the events. See also Appendix \ref{AppA}.

We  remark that the sum over single-history amplitudes is made possible by the assumption of internal causality. At the level of causets, the histories still can intersect. Given a subset $A'$ of a complete acausal set $A$,  the number of events in $A'$ will change amidst evolutions. To see this,  we can introduce two operations on events in the acausal set $A'$: one is the coarse-graining $\mathfrak{C}$ of  events; the other is the fine-graining $\mathfrak{F}$ of events. For the example of two events $p_1,p_2\in A'$, these act as
\begin{equation}\label{CF}
\mathfrak{C}:\{p_1,p_2\}\mapsto \{p_1=p_2\},\quad \mathfrak{F}:\{p_1\}\mapsto\{p_3,p_4\}
\end{equation}
where the events $p_3,p_4\in A$ are such that  $A'\cup\{p_3,p_4\}/\{p_1\}$ is still acausal.  Here the causal evolutions have been encoded in causal histories, and these operations $\mathfrak{C}$ and $\mathfrak{F}$ can be merely interpreted as coarse-grainings and fine-grainings of the acausal set making the histories to intersect or to branch. 

 We hope that the total amplitude is invariant, or cylindrically consistent, under the actions  of $\mathfrak{C}$ and $\mathfrak{F}$.
For the coarse-grainings,  the single-histories $\mathsf{h}_\gamma$ are unaltered if one recovers the events before a coarse-graining by internal causality. But the (spaces of) states will change after these operations, since the events on which the states are assigned are changed.
 In this sense, the operations  effectuate the transitions between different single-history states $\ket{\mathsf{h}_\gamma}$ on the same history $\mathsf{h}_\gamma$ but with different states. In other words, these operations induce {\it fluctuations} of states around the history $\mathsf{h}_\gamma$. We can take the reference states $\ket{\psi_i}_{\text{ref}}$ assigned to a quantum causal history $\mathsf{h}_\gamma$ as its ``stationary" states, then the fluctuations around  $\ket{\psi_i}_{\text{ref}}$ at each event is encoded in the tensorial coefficients $M^{(i)}$ as $\sum_iM^{(i)}\ket{\psi_i}_{\text{ref}}$.
In this way, the single-history MPS $\ket{\mathsf{h}_\gamma}$ can be understood as the superposition of all fluctuations around $\mathsf{h}_\gamma$, as a consequence of which the sum over histories \eqref{16} indeed sums over all these fluctuations.
As for the fine-grainings, generically there will  be new events (and hence new histories) added to $A'$, in addition to the changes in the states. The addition of new events amounts to the fluctuations in the quantum causal history $\mathsf{h}_\gamma$ itself. In the sum over histories \eqref{16}, since the trajectories $\gamma$'s are not fixed, we can simply add some new trajectories.
In this sense,  the total amplitude of the sum over histories \eqref{16} sums over all the fluctuations of the histories and is therefore cylindrically consistent.

 Notice that we can also interpret these operations  $\mathfrak{C}$ and $\mathfrak{F}$ as local time evolution moves \cite{DS14} of the acausal set $A'$. In this interpretation we also need to define the identity evolution moves for events  upon which the $\mathfrak{C}$ and $\mathfrak{F}$ do not act:
\begin{equation}
\mathfrak{I}: \{p_j\in A'|j\in J\}_t\mapsto \{q_j\in A''|j\in J\}_{t+1}
\end{equation}
where $J$ is the index set  for events in the kernels of $\mathfrak{C}$ and $\mathfrak{F}$. $\mathfrak{I}$ is a  one-to-one map keeping the number of events. Again by internal causality, these evolution moves $\mathfrak{C},\mathfrak{F}$ and $\mathfrak{I}$ have been encoded in the MPS of single-histories.

\subsection{Entangled quantum causal histories}\label{3.2}
We have constructed the MPS of a single quantum causal history from one-dimensional PEPS. It is natural to ask what the history states constructed from higher dimensional PEPS are. Starting with single-history states, one expects the possible two-dimensional PEPS to  be the entangled superpositions of these single-history states. In \cite{CW16}, the entangled histories are studied  for consistent histories. Here we shall consider entangled quantum causal histories.\footnote{In  \cite{CW16}, the  single histories are defined on distinct spatial quantum trajectories, so by superposing history states on distinct trajectories one can construct entangled histories. But for a single quantum causal history, the underlying trajectory is unique and its MPS is really an entangled state on a single trajectory. A simple way to see this is that for consistent histories the projections in a class operator are into the space coordinate eigenstate, e.g. $\ket{x_i}\bra{x_i}$, while for quantum causal histories, the spacetime metric has not been constructed yet and the states residing on event are only internal for now.
}

 Intuitively, given a set of states $\ket{\mathsf{h}_\gamma}$ of single histories, one can get a general superposed history state $\ket{\mathsf{H}}=\sum_\gamma a_\gamma\ket{\mathsf{h}_\gamma}$ with $a_\gamma\in\mathbb{C}$. To formulate   inner products for the superposed history states $\ket{\mathsf{H}}$, we observe  an obvious difference between consistent and quantum causal histories: in defining the single-history state $\ket{\mathsf{h}}$ the causal evolutions between events have been taken into account, while in consistent histories one needs the class operators to specify particular histories. As a consequence, 
the inner product of two single-history states depends solely on the relations between two quantum causal histories.
For example, for disjoint histories $\mathsf{h}_{\gamma},\mathsf{h}_{\gamma'}$, we should have $\braket{\mathsf{h}_{\gamma}|\mathsf{h}_{\gamma'}}=0$. With the assumption of internal causality at events, we only need to consider disjoint quantum causal histories with internal causality, so that a non-vanishing inner product is defined only when two quantum causal histories are consistent, i.e. $\braket{\mathsf{h}_{\gamma}|\mathsf{h}_{\gamma}}\neq0$. In this way, an inner product of $\ket{\mathsf{H}}$'s can be defined as usual without referring to class operators.

Formally, one can consider the normalized superposed states $\ket{\bar{\mathsf{H}}}$ and define a ``higher-level" complete orthonormal basis  $\{\ket{\bar{\mathsf{H}}_i}\}$. That is, one requires that $\braket{\bar{\mathsf{H}}_i|\bar{\mathsf{H}}_j}=0$ for $i\neq j$ and $\sum_ib_i\ket{\bar{\mathsf{H}}}$ gives a complete set of histories. One can also construct ``higher-level" entangled states based on the basis $\{\ket{\bar{\mathsf{H}}_i}\}$. However, unlike the consistent histories, the causal evolutions are now CPTP maps whose behaviors under linear superposition is not easy to see. Moreover, superposing different quantum causal histories will mix the definite causal relations in these histories, leading to indefinite causal structures. Although there are attempts to study quantum gravity with indefinite causal structure \cite{Har05}, these theories are drastically different from the causet theory we have used here.\footnote{The most obvious difference is the treatment of causality: in causet the causal relations of each history are definite and one actually only knows the causal relations at the outset, whereas an indefinite causal structure makes the causal order between events unfixed. In addition, causets are pre-geometric theories, while the existing theories with indefinite causal structure are operational theories with much wider applicability.}
To avoid further conceptual difficulties, we want to define superposed or entangled quantum causal histories with their respective causal relations kept intact. 
An apparent example is a two-arm single-particle interferometer where the two quantum states on the two arms can be entangled but their causal evolutions along the arms are definite. Let $H_1$ be the Hilbert space of states localized in the first arm, and $H_2$ in the second arm, then the Hilbert space of a single particle in the two-arm interferometer is $H=H_1\oplus H_2$. If on each arm the evolution is determined by some CP (not necessarily TP) map,  the  superpositions of two CP maps on the two arms have been deeply studied by \AA berg \cite{Abe04}. What we need from \cite{Abe04} is the following:

Let $\varphi:B(H_S)\rightarrow B(H_T)$ be a CP map from the  algebra of bounded linear operators on the  source ($S$) Hilbert space to that on the target ($T$) Hilbert space.\footnote{Here the direction of the map is treated for simplicity. The relation to that in Def. \ref{d21} is obvious.}
 Suppose $H_S$ and $H_T$ are all finite-dimensional.
 Consider the case in which the source and target Hilbert spaces are respectively decomposed into orthogonal sums, $H_S=H_{s1}\oplus H_{s2}$ and $H_T=H_{t1}\oplus H_{t2}$. Denote by $P_{si},P_{tj},i,j=1,2$ the projection operators onto the respective subspaces, then for any operator $Q\in B(H_S)$, the action of a CP map $\varphi$ is 
\begin{equation}
\varphi(Q)=\sum_{i,j,k,l=1,2}P_{ti}\varphi(P_{sj}QP_{sk})P_{tl}.
\end{equation}
When $\varphi$ is a CPTP map, we say $\varphi$ is {\it subspace preserving} if 
\begin{equation}
\text{tr}(P_{ti}\varphi(Q))=\text{tr}(P_{si}Q),\quad i=1,2,\quad\forall Q\in B(H_S)
\end{equation}
which means that the probability weight on each subspace in the orthogonal sum decomposition (on the arm of the interferometer) is preserved under the CPTP map. Importantly, a CPTP map $\varphi$ is subspace preserving iff it is a {\it ``gluing"} of two CPTP maps $\varphi_1,\varphi_2$ defined on the subspaces. Here by ``gluing" we mean that when a CP map $\varphi$ is restricted to the source subspace $H_{s1}$ (or $H_{s_2}$) and to the target subspace $H_{t1}$ (or $H_{t2}$), the restriction $\varphi|_1\equiv \varphi_1$ (or $\varphi_2$) is also a CP map on the subspace $H_{s1}$ (or $H_{t2}$), to wit
\begin{equation}
\varphi|_{H_{si},H_{ti}}(Q)=P_{ti}\varphi(P_{si}QP_{si})P_{ti}=\varphi_i(Q_i),\quad i=1,2,
\end{equation}
where $Q_1\in B(H_{s1}),Q_2\in B(H_{s_2})$ and $Q\in B(H_S)$. The ``gluing" of CPTP maps is in effect the superposition of two evolutions on those two arms in the interferometry setup.

The ``gluing" of CPTP maps is indeed a  suitable concept for superposing quantum causal histories. On the one hand, the orthogonal sum of two Hilbert spaces allows general quantum coherence including entanglement between states of the two spaces. On the other hand, since the subspace preserving property preserves the probability weights under joint evolutions, the causal relations in a single  history, which present themselves as the transition probabilities in a (quantum) Markov process, are preserved under joint evolutions. In this way, the indefinite causal structure for entnagled quantum channels is avoided. 

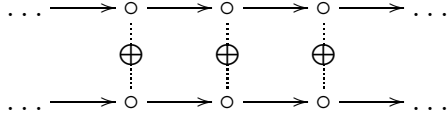
\begin{figure}[t]
\[
\xymatrix{
\dots\ar[r] & {\circ}\ar@{.}|{\bigoplus}[d]\ar[r]  & {\circ}\ar@{.}|{\bigoplus}[d]\ar[r] & {\circ}\ar@{.}|{\bigoplus}[d]\ar[r]&\dots\\
\dots
\ar [r] 
& {\circ}\ar[r] & {\circ}\ar[r] & {\circ}\ar[r] &\dots
 }
\]
\caption{Schematic ``gluing" of quantum causal histories. }\label{f33}
\end{figure}

To be definite, we take the entangled  two events, which are possibly the same but with internal causality, of two  distinct quantum causal histories to be in the same acausal set $A$.
As an example, consider two events $p_1,p_2\in A$, through which passes two quantum causal histories $\mathsf{h}_1,\mathsf{h}_2$. At $p_1$ the projection operator $\beta_{p_1}$ of the single history $\mathsf{h}_1$ should be modified by a new part coming from the entangled pair between states at $p_1$ and $p_2$. Let us write
\begin{equation}\label{2222}
\beta_{p_1}=\sum_{ijkl}M^{(p_1)}_{ijkl}\ket{\psi_k^{(p_1)}}\hat\phi^\dag(\bra{\psi_{P_1-1,i}})\bra{\psi_{p_1,j}}\bra{\xi_{p_1,l}},
\end{equation}
and similarly at $p_2$, where $\sum_l\ket{\xi_{p_1,l}}\ket{\xi_{p_2,l}}$ is the entangled pair shared by $p_1$ and $p_2$. Obviously such projection operators project out the states from the inter-history entangled pairs if we restrict ourselves to $\mathsf{h}_1$, thereby preserving the causal relations as well as the causal evolutions in $\mathsf{h}_1$. But for joint evolutions of $\mathsf{h}_1,\mathsf{h}_2$, the ``gluing" of subspace states have the form of
\begin{equation}
\sum_{i_1i_2...i_n}\bigl(a_1\ket{\mathsf{h}_1}\ket{\xi_{1,i_1i_2...i_n}}+a_2\ket{\mathsf{h}_2}\ket{\xi_{2,i_1i_2...i_n}}\bigr),\quad a_1,a_2\in\mathbb{C}
\end{equation}
and therefore the subspace evolutions can be regained by using the projection operators $P_{\xi_i}=\sum_j\ket{\xi_{i,j}}\bra{\xi_{i,j}},i=1,2$. The Hilbert spaces $H_{p_1},H_{p_2}$ on the events $p_1,p_2$ are thus ``glued" to $H_{p_1}\oplus H_{p_2}$.
Denoting the Hilbert space of the single-history states $\ket{\mathsf{h}_i}$ by $\mathcal{H}_i$, we obtain the direct or orthogonal sum $\mathcal{H}_1\oplus\mathcal{H}_2$. See for example Figure \ref{f33}.

As for the tensor network representation of such ``gluing" of histories, we only need to apply the projection operators as in \eqref{2222} at each event to perform the higher dimensional PEPS construction. On the other hand, we have at each node $v$ of the underlying graph of the joint histories                                                     
\begin{equation}\label{VV}
\ket{M_v}=\sum_{ijkl}V_{ijkl}^{(v)}\ket{\psi_i}\ket{\psi_j}\ket{\psi_k^{(v)}}\ket{\xi_l^{(v)}}
\end{equation}
where $\ket{\xi_l^{(v)}}$ is part of the (maximally) entangled pair $\sum_{l}\ket{\xi_l^{(v)}}\ket{\xi_l^{(v')}}$ with $v,v'$ being the nodes on different histories but in the same complete acausal set. In addition to the  bipartite bond states, there are also inter-history entangled bond states 
\begin{equation}\label{EE}
\bra{\xi_v}=\sum_{i}\bra{\xi_i^{(v)}}\bra{\xi_i^{(v')}}.
\end{equation}
We can  formulate the following formal definition.
\begin{construction}\label{3311}
Let  $\mathsf{h}_i,i=1,2$, be two quantum causal histories.
The tensor network representation of the ``gluing" of two histories  is 
\begin{equation}
\ket{\mathsf{h}_1\oplus\mathsf{h}_2}=\bigotimes_e\bra{\phi_e}\bigotimes_v\bra{\xi_v}\bigotimes_v\ket{M_v}
\end{equation}
where the  $\ket{M_v}$ are given by \eqref{VV}, the $\bra{\phi_e}$ by \eqref{E}, and the $\bra{\xi_v}$ by \eqref{EE}.
\end{construction}
Two remarks are in order. Firstly, this kind of ``gluing" of histories is a ``second-level" entangled structure in the sense that both in the history direction and in the inter-history direction we have entangled states. But it is not an isotropic PEPS even if one consider superposes $n$ histories, because in the inter-history direction there is no causal evolutions. 
Secondly,
 the choice of the entangled pairs is not unique, and it is event possible to consider only quantum coherences between two histories. The choice of maximally entangled pairs in \eqref{EE} is again motivated by the results in loop quantum gravity. The kinematical Hilbert space of loop quantum gravity is $H_{\text{kin}}=\bigoplus_e\bigotimes_v H_{v,e}$, where $v$ and $e$ are vertices and edges of a spin network graph. From this we can see that the vertex Hilbert spaces are glued by colored edges, which are required to carry maximally entangled states of intertwiners \cite{BBY18}. If the inter-history entanglement exists only for events in the same {\it complete} acausal set $A$, the ``gluings" of quantum causal histories induce the direct sum of Hilbert spaces on events of $A$, which is exactly the structure of the kinematical Hilbert space for the spin network states in a three-dimensional spacelike slice. In this case, the ``gluings" of events in causet and the gluings of polyhedra in loop quantum gravity then nicely match.

\subsection{Towards  holographic quantum causal histories}\label{3.3}
In addition to MPS and PEPS, another typical tensor network is MERA which can have a holographic description \cite{holography}. We try to find a holographic relation in the tensor network representation of quantum causal histories.  Possible holographic properties in quantum causal histories have been explored in \cite{MS99} where on each event is assigned a ``holographic" screen and the causal evolutions between events are assumed to be the flows of quantum information between screens.  Instead of the bulk-boundary holography, this kind of holography is a local one without referring to the bulk theory. Besides, the interpretation of quantum information flow only applies for the locally unitary evolutions, otherwise there will be information loss, so this assumption cannot be directly imposed on CPTP maps. A nice way to amend these problems is to supplement a quantum channel $\varphi$ with a local environment but keep the causal structure, so that the unitary evolution of the purified history-environment system can be implemented by some local isometries. Using these isometries one can construct a de Sitter MERA \cite{Ben13}.
Here we first formulate a coarse-graining mapping analogous to the unitary {\it exact holographic mapping} \cite{Qi13}, but it is not exactly holographic since it does not preserve all the information. Then we argue that the { holographic purification}  is a possible way to implement the local history-environment total unitary system, thereby allowing the  MERA construction on quantum causal histories.

\begin{figure}[t]
 \[\xymatrix{&p_1&& p_2  &\\
 &  & q \ar[ru]\ar[lu] & &\Longrightarrow\\
&r_1\ar[ru]&&   r_2\ar[lu]  &  
}
\xymatrix{&p_1&& p_2 \\
 &  & q \ar[ru]\ar[lu] & \\
&r_1\ar[ru]&&   r_2\ar@{--}[lu]  
}\]
\caption{Holographic mapping for quantum causal histories.}\label{f3}
\end{figure}
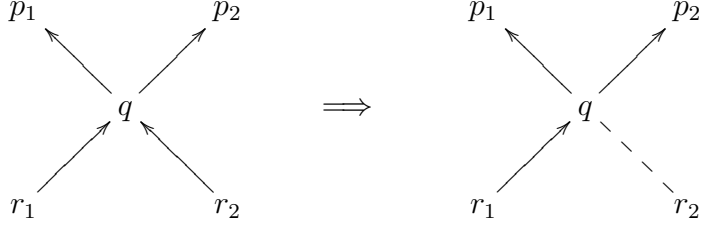

Consider an acausal set $A$ of spacelike separated events $p_1,p_2,...,p_n$. Recalling from Def. \ref{d21} that a CPTP map $\phi_{qp_1}:\mathcal{U}(p_1)\rightarrow\mathcal{U}(q)$ maps the type I factor $\mathcal{U}(p_1)$ on $p_1$ into  $\mathcal{U}(q)$ on some event $q$ in its causal past, we can take the fine-grainings in the direction of causal evolution as the coarse-grainings in the direction of the actions of CPTP $\phi$'s.
In the above, we have assumed the internal causality for those quantum causal histories coinciding in the same event. Let $\mathcal{U}(p_1),\mathcal{U}(p_2)$ be the algebras on $p_1,p_2\in A$ respectively, then the CPTP maps $\phi_{p_1q}:\mathcal{U}(p_1)\rightarrow\mathcal{U}(q)$ and $\phi_{p_2q}:\mathcal{U}(p_1)\rightarrow\mathcal{U}(q)$ both end on the algebra $\mathcal{U}(q)$ over the event $q\in P(A)$. 
With the assumption of internal causality, there would be events $r_1,r_2$ such that the two quantum causal histories $\{r_1\leqslant q\leqslant p_1\}$ and $\{r_2\leqslant q\leqslant p_2\}$ are effectively disjoint with e.g. $q_1\curlyeqprec q_2$. At the event $q$, there are also two projection operators $\beta_{q_1},\beta_{q_2}$, realizing the PEPS construction,
\begin{equation}
\beta_{q_1}=\sum_{ijk}M^{(q_1)}_{ijk}\ket{\psi_k^{(q_1)}}\hat\phi^\dag(\bra{\psi_{r_1,i}})\bra{\psi_{p_1,j}},\quad 
\beta_{q_2}=\sum_{ijk}M^{(q_2)}_{ijk}\ket{\psi_k^{(q_2)}}\hat\phi^\dag(\bra{\psi_{r_2,i}})\bra{\psi_{p_2,j}}.
\end{equation}
Now let us relax the internal causality (letting $q_1=q_2$), and discard in the action of $\beta_{q_2}$ the transformation from $\check\phi^\dag(\ket{\psi_{r_2,i}})$ to the reference state $\ket{\psi_{p_2,j}}$. Then $\check\phi^\dag(\ket{\psi_{r_2,i}})$  does not qualify for the next-step causal evolution in our constructions. We can still define the tensor networks by basing   the two events $p_1,p_2$ on a common  past $\{r_1\leqslant q\}$. The projection $\beta_{p_2}$ is changed to
\begin{equation}\label{311}
\beta'_{q}=\sum_{ijk}M^{\prime(q)}_{ijk}\check\phi^\dag(\ket{\psi_{r_2,i}})\hat\phi^\dag(\bra{\psi_{r_1,i}})\bra{\psi_{p_2,j}}
\end{equation}
with the tensorial coefficients $M'$ encoding the transformations from $\hat\phi^\dag(\bra{\psi_{r_1,i}})$ to $\bra{\psi_{p_2,j}}$. The states $\check\phi^\dag(\ket{\psi_{r_2,i}})$ evolved from the event $r_2$ are now left out in the tensor network construction, which, however, constitute  the {\it bulk states} living on bulk events, while the remaining tensor networks, in which each history is a MPS, provide a fine-graining flow in the direction of causal evolution. See for instance Figure \ref{f3}.
 In the direction of CPTP maps, we have a coarse-graining flow of the algebras on events that maps out a two-dimensional  net of algebras from the algebras on the acausal set $A$.

\begin{figure}[t]
\[
\xymatrix{
\dots\ar[r] & {\circ}\ar[r]  & {\circ}_p\ar@{=>}[d]\ar[r] & {\circ}\ar[r]&\dots\\
\dots
\ar '[r] 
'[rr]+D*{\bullet} 
& {\circ}\ar[r] & {\circ} &  &
 }
\]
\caption{Schematic holographic purification. The boundary is at $p$, and the causal future is compressed into a state (black dot) near the boundary. The total system (black+white dots) is a pure state following unitary evolutions. }\label{f4}
\end{figure}
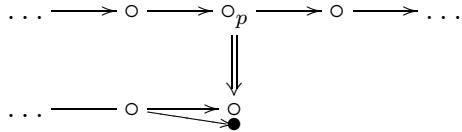

The above flows are constructed in a way similar to the exact holographic mapping, but with the obvious difference that the CPTP maps are not unitary and even not isometries. Due to these non-unitary local causal evolutions, many results from the tensor networks with local unitaries cannot be directly applied here. But recall that a quantum state that can be approximated by a MPS usually follows an area law. One can indeed formulate an area law for a single history by bounding the tensorial coefficients in the MPS.
Let us suppose that a single-history state, in addition to be a MPS, is a  one-dimensional gapped quantum system following an area law. This means that if we cut the one-dimensional quantum causal history at an event $p$, the von Neumann entropy $S(F(p))$ of the causal future $F(p)$ is bounded as $S(F(p))\leqslant k$ for some constant $k$. Then by the  {\it holographic compression theorem} \cite{WE19}, the state $\ket{F(p)}$ on $F(p)$ can be unitarily compressed into a state $\ket{N(p)}$ near the boundary event $p$ up to some error $\epsilon$ depending on $k$. If we denote by $\ket{\mathsf{h}_{P(p)}}$ the single-history state  on the causal past of $p$,  the joint state $\ket{\mathsf{h}_{P(p)}}\otimes\ket{N(p)}$ can be made into a holographically purified state. See also Figure \ref{f4}.
The causal future of $p$ can be unfolded by acting a unitary on the purified state, so that the causal evolutions in a single causal history can be holographically  understood as the unitary evolutions of the purified states. 

Another way to understand this is to note that the compressed state $\ket{N(p)}$ is near the boundary event $p$ up to some error $\epsilon$. As a result, the correlation functions of any operators in these states have almost-vanishing correlation lengths $\sim\epsilon$. In other words, the compressed states are renormalization group {\it fixed point} states, where the renormalization group is provided by the tensor network renormalization group defined for example  by  the holographic mapping \eqref{311}. The addition of future events will all be compressed into the states $\ket{N(p)}$ and will not affect the tensor network states defined up to $p$.  This also in some sense assures the scale invariance of a MERA.

With these local purifications,
one can construct a de Sitter MERA along the  lines of \cite{Ben13}, or use the holographic mapping suggested above.

\section{Example from  temporally varying discretization }\label{sec4}
In this section, we discuss an example of the tensor network representation for local quantum causal histories in the theory of fully constrained quantum systems on  temporally varying discretizations \cite{Hoh14}. This kind of temporally varying discretizations are naturally  associated to discretization-changing dynamics such as those
 in the canonical simplicial gravity and cosmological mode creations. The non-unitary maps are shown to be necessary for defining a local causal structure in temporally varying discretizations, and they also give a simple representation of the black hole horizon in the related tensor networks.
\subsection{Re-interpretation of temporally varying discretization }
Consider a quantum system defined on a temporally varying discretization of spacetime. To be concrete, let the discretized space be a graph $\Gamma$, and suppose the time evolution moves of $\Gamma$ are between discrete time steps. Global time evolution moves evolve the entire graph $\Gamma_n$ at the $n$-th time step into the graph $\Gamma_{n+1}$ at the next time step, which is generically graph-changing. 
Let $H^{\text{kin}}_n$ be the kinematical Hilbert space assigned to $\Gamma_n$, then $\dim H_n^{\text{kin}}\neq\dim H_{n+1}^{\text{kin}}$ when the time evolution moves are graph-changing.

If  the system is furthermore a totally constrained system with only first class constraints (with the second class constraints solved at the classically), there exist (Hamiltonian) constraint operators ${C}_{n,I}$ at the $n$-th step such that the physical states of $\Gamma_n$ are selected by the group averaging projector, or rigging map,
\begin{equation}
P_n=\prod_I\delta({C}_{n,I}):~H_n^{\text{kin}}\rightarrow H_{n}^{\text{phys}};~\ket{\psi_n^{\text{kin}}}\mapsto P_n\ket{\psi_n^{\text{kin}}}=\ket{\psi^{\text{phys}}_n}.
\end{equation}
Denoting by $x_n$ the  configurations of $\Gamma_n$, we can express the  evolution of the  state  through some propagators $K(x_{n+1},x_n)$ as
\begin{equation}\label{333}
\ket{\psi_{n+1}^{\text{phys}}(x_{n+1})}=\int dx_nK(x_{n+1},x_n)\ket{\psi_n^{\text{kin}}(x_n)}
\end{equation}
if the dimension the kinematical Hilbert space does not change.
The group averaging projectors $P_n,P_{n+1}$ have been incorporated into $K(x_{n+1},x_n)=P_{n+1}(P_n)^*k(x_{n+1},x_n)$ where $k(x_{n+1},x_n)$ is the propagator on the kinemtatical Hilbert spaces. 

When the evolution moves are graph-changing, the dimensions of the kinematical  Hilbert spaces will change, e.g. $\dim H_n\neq\dim H_{n+1}$. We can extend the kinematical Hilbert spaces  $H_n^{\text{kin}},H_{n+1}^{\text{kin}}$ according to the evolution step $n\rightarrow n+1$ such that $\dim \bar{H}_n=\dim \bar{H}_{n+1}$ as follows:
\begin{enumerate}
\item extend $H_{n}^{\text{kin}}$ by adding to it ``new" configurations $x_{\mathfrak{n}}$ that appear in $H_{n+1}^{\text{kin}}$ but are absent in $H_{n}^{\text{kin}}$, resulting $\bar{H}_{n}^{\text{kin}}$;
\item extend $H_{n+1}^{\text{kin}}$ by adding to it ``old" configurations $x_{\mathfrak{o}}$ that appear in $H_{n}^{\text{kin}}$ but are absent in $H_{n+1}^{\text{kin}}$, with the resulting extended Hilbert space $\bar{H}_{n+1}^{\text{kin}}$.
\end{enumerate}
These extensions are required to be cylindrically consistent such that the physical states (and the physical  inner products) are kept intact. In this way, we can still define the kinematical propagators $\bar{k}(x_{n+1},x_n)$ on the extended Hilbert spaces.

\begin{figure}[t]
\[
\xymatrix{
x_{k}^{1} \ar@{-}[dr] &&x_{k}^2 \ar@{-}[dl]&~\ar @{} [dl] |{\Longrightarrow}\\
 & x_k^\mathfrak{o} & }
\xymatrix{
x_{k+1}^{1}\ar@{-}[rr] &&
x_{k+1}^2 }
\]
\caption{A $2-1$ Pachner move.}\label{fP}
\end{figure}
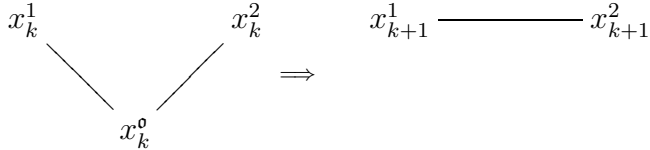

By \eqref{333} the global physical states at each time step can be obtained from the initial kinematical state. Now suppose we know the global physical state $\ket{\psi_k^{\text{phys}}}$ after $k$ steps of  evolution, and consider local evolution moves of $\Gamma_k$. Here by local we mean only some nodes in a local region $\gamma_k$ of $\Gamma$ is changed by the time evolution moves. These local moves will result in a new global physical state  $\ket{\psi_{k+1}^{\text{phys}}}$ which differs from $\ket{\psi_k^{\text{phys}}}$ only by these local moves. By \eqref{333} we can construct a physical state updating transform $\mathfrak{u}_{k\rightarrow k+1}$ between  $\ket{\psi_{k}^{\text{phys}}}$ and $\ket{\psi_{k+1}^{\text{phys}}}$, 
\begin{equation}
\ket{\psi_{k+1}^{\text{phys}}}=\mathfrak{u}_{k\rightarrow k+1}(\ket{\psi_{k}^{\text{phys}}}).
\end{equation}
What  $\mathfrak{u}_{k\rightarrow k+1}$ interests us is that for some cases it is not unitary \cite{Hoh14}. Let us consider the simple case of $2-1$ Pachner moves (e.g. Figure \ref{fP}). 
It is not difficult to see that not all of the physical state $\ket{\psi_k^{\text{phys}}}$ will evolve to the next step $k+1$, because the configuration $x_k^\mathfrak{o}$ is moved. We need a new constraint operator $C'_k$ such that the corresponding group averaging projector $P'_k$ selects the evolving part $P'_k\ket{\psi_k^{\text{phys}}}\neq0$. It is proved in \cite{Hoh14} that the physical inner product are not preserved,
\begin{equation}
\braket{\psi_{k+1}^{\text{phys}}|\phi_{k+1}^{\text{phys}}}=\braket{\psi_{k}^{\text{phys}}|P'_k|\phi_{k}^{\text{phys}}}\neq \braket{\psi_{k}^{\text{phys}}|\phi_{k}^{\text{phys}}}.
\end{equation}
If  the Dirac observables at the step $k$ all commute with the constraint $C'_k$, the evolution $k\rightarrow k+1$ is still unitary. But if not, the $2-1$ Pachner moves can be non-unitary.

The above non-unitary evolution moves can be rewritten in the formalism of quantum causal histories.  Indeed, we can take the configurations $x_k^i$, e.g. coordinates, at the nodes of $\Gamma_k$ as the events in a causet, and the time evolution moves as causal evolutions. The links in $\Gamma_k$ are other relations between events than the causal ones, so that $\Gamma_k$ is a spatial Cauchy surface or a complete acausal set. The quantum states $\ket{\psi^{\text{kin}}(x_k^i)}$ are then quantum states on events, and the global state is a compound state of these local states, which can be superposed or just product states.
The distinction between kinematical and physical states allows us to select a physical subset $A^{\text{phys}}$ of the acausal set $A$.  
Then as in \cite{DS14}, the global or local evolution moves of $A^{\text{phys}}$ can be interpreted as the operations of coarse-graining $\mathfrak{C}$, fine-graining $\mathfrak{F}$, entangling $\oplus$ and the identity $\mathfrak{I}$, all of which have been introduced in Sec. \ref{sec3}. When composing these evolution moves, one needs to match the constraint operators of different steps, or to add new constraints if non-unitary evolution moves are possible, which corresponds to the projection in the PEPS construction. However, for the unitary evolution moves, such a rewriting is rewardless, so we focus on the non-unitary part:

In the above example of $2-1$ Pachner move, we can alternatively take the configurations $(x_k^1,x_k^\mathfrak{o},x_k^2)$ as an event $e$.
These configurations $(x_k^1,x_k^\mathfrak{o},x_k^2)$ in $e$ are acted upon by $\mathfrak{F}$, while the rest configurations in the extended kinematical Hilbert space $\bar{H}_k^{\text{kin}}$ remain unchanged or undergo the action of $\mathfrak{I}$. More generally we can consider the configurations $(x_k^\mathfrak{o},x_k)$ of $e$ that undergo non-unitary local evolution moves.
 Since only these configurations of $e$ are changing, the physical states $\ket{\psi^{\text{phys}}_k}$ at the step $k$ can be chosen to be the states $\ket{\psi^e_k}$ localized on $e$. These states $\ket{\psi^e_k}$ can be  obtained by   tracing over all the states acted by $\mathfrak{I}$, i.e. \begin{equation}
\rho_k^e=\text{tr}_\mathfrak{I}\ket{\psi^{\text{kin}}_k}\bra{\psi^{\text{kin}}_k},\quad \ket{\psi^{\text{kin}}_k}\in\bar{H}_k^{\text{kin}}.
\end{equation}
 Now the physical state updating map $\mathfrak{u}_{k\rightarrow k+1}:\rho^e_k\rightarrow\rho_{k+1}^e$ is simply the mapping of the states over the event $e$. When $\mathfrak{u}_{k\rightarrow k+1}$ is non-unitary, it becomes a map between the density matrices, which is a CPTP map.\footnote{The trace preserving property is obvious from the operation $\mathfrak{I}$. The complete positivity is assured by the freedom in extending the kinematical Hilbert spaces.}
We thus obtain a  single quantum causal history $\mathsf{h}_e$ passing through the event $e$ where each causal evolution is described by a CPTP map of physical state updating.

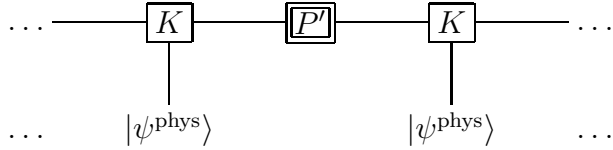
\begin{figure}[t]
\[
\xymatrix{
\dots\ar@{-}[r] & *+[F]{K} \ar@{-}[d]\ar@{-}[r] &  *+[F=]{P'}\ar@{-}[r] &  *+[F]{K}\ar@{-}[d]\ar@{-}[r]&\dots\\ 
\dots& \ket{\psi^{\text{phys}}} &  & \ket{\psi^{\text{phys}}} &\dots
 }
\]
\caption{The history state of  local moves. }\label{fTVD}
\end{figure}

The structure of totally constrained system allows us to construct tensor networks without using the channel-state duality.
Let us recall that for a single step of local evolution move $k\rightarrow k+1$,  the  constraint operators at step $k$ and step $k+1$ are classified into three classes \cite{Hoh14}: the first class of constraints are those preserved by the evolution move, i.e. $C_k=C_{k+1}$; the second  are the new constraints $C'_{k}$ added to $H_k$, as in the example of $2-1$ Pachner move; similarly the third are the new constraints $C'_{k+1}$ added to $H_{k+1}$. Then the physical state updating map $\mathfrak{u}_{k\rightarrow k+1}$ can be, in analogy to \eqref{333}, expressed by a kinematical propagator $u(x_{k+1},x_k) $ as
\begin{equation}\label{337}
\ket{\psi^{\text{phys}}_{k+1}}=\int dx_k^{\mathfrak{o}} P'_{k+1} P^*_{k/k+1}P^{\prime *}_k u(x_{k+1},x_k)  \ket{\psi_k^{\text{phys}}}
\end{equation}
where the integration is over the ``old" configurations $x_k^{\mathfrak{o}}$, and the $P$'s are the group averaging projectors corresponding to the three classes of constraint operators. On the other hand,  we can obtain each physical state from an initial kinematical state $\ket{\psi_0^{\text{kin}}}$ by exploiting \eqref{333},
\begin{equation}\label{338}
\ket{\psi_{k}^{\text{phys}}}=\int dx_0P_{k}P_0^*k(x_{k},x_0)\ket{\psi_0^{\text{kin}}}.
\end{equation}
By reflecting  on \eqref{337} and \eqref{338}, we can choose the reference state of the tensor network construction as the initial kinematical state $\ket{\psi_0^{\text{kin}}}$, and redefine the physics initial state of the history to be $\ket{\psi_{1}^{\text{phys}}}$ at step $1$.
In other words, we can place a $\ket{\psi_0^{\text{kin}}}$ on each event in $\mathsf{h}_e$, and then perform the propagator transformation \eqref{338} to get the physical state $\ket{\psi_{k}^{\text{phys}}}$ on the event at step $k$. Then the physical state $\ket{\psi_{k}^{\text{phys}}}$ at step $k$ is mapped to the physical state $\ket{\psi_{k+1}^{\text{phys}}}$ at step $k+1$ via the physical state updating transformation \eqref{338}. Schematically,  such a history state is depicted in Figure \ref{fTVD}, where the $K$'s on nodes represent the propagator transformation \eqref{338} and the $P'$ on links represent the newly added  constraints. Two $K$'s can be glued if the constraint operators of them match and the added constraints are satisfied.

Based on Figure \ref{fTVD}, we can return to the description via CPTP maps, and apply the construction of Sec. \ref{sec3}. The resulting MPS is similar to the single-history state \eqref{10}, with the residing states being the physical states. To extend this single-history MPS to the case with global moves, one can use the operations $\mathfrak{C},\mathfrak{F},\mathfrak{I}$ and $\oplus$ to multiple histories. But these operations do not cover all the known Pachner moves. So further discussions require more  tensor network constructions.

\subsection{``Light cone" and quantum black hole}
Consider two operators $O_{e_1}$ and $O_{e_2}$  on the events $e_1$ and $e_2$ respectively. If these two events $e_1$ and $e_2$ in $\Gamma_k$ are ``spacelike" separated in the sense of Einstein locality, the operators satisfy $[O_{e_1},O_{e_2}]=0$, which also holds when measured in the physical states on $\Gamma_k$ that includes $e_2$ and $e_2$:
\begin{equation}\label{oo}
\bra{\psi^{\text{phys}}_k}[O_{e_1},O_{e_2}]\ket{\psi^{\text{phys}}_k}=0.
\end{equation}
Here the $O_{e_1}$ and $O_{e_2}$ are assumed to have non-vanishing eigenvalues when acting on all the physical states.

The $\ket{\psi^{\text{phys}}_k}$ will undergo the graph-changing local evolution moves, and the locality of events can be changed by the non-unitary moves. Then after $n$ steps of local moves with some of them non-unitary, the expectation value \eqref{oo} will become non-vanishing if the $e_1$ or $e_2$ are no longer separated. A ``light cone" structure, say on $e_1$, is thus given by the minimal $n$ such that the $e_1$ or $e_2$ are still separated:
\begin{equation}
\bra{\psi^{\text{phys}}_{k+n}}[O_{e_1},O_{e_2}]\ket{\psi^{\text{phys}}_{k+n}}=0.
\end{equation}
This structure can only be defined if there are non-unitary moves, because the unitary moves preserve the Dirac observables, i.e. $[O,C'_k]=0$, so that
\begin{equation}
\bra{\psi^{\text{phys}}_{k+1}}[O_{e_1},O_{e_2}]\ket{\psi^{\text{phys}}_{k+1}}=\bra{\psi^{\text{phys}}_{k}}P_k^{\prime\dag}[O_{e_1},O_{e_2}]P'_k\ket{\psi^{\text{phys}}_{k}}=\bra{\psi^{\text{phys}}_{k}}[O_{e_1},O_{e_2}]\ket{\psi^{\text{phys}}_{k}}.
\end{equation}
For non-unitary moves, we may have $[O_{e_2},C'_k]\neq0$, and hence the expectation value \eqref{oo} will change.

This ``light-cone" structure is in fact a Schr\"odinger-picture version of
 the bulk causal structure defined by the generalized butterfly velocity in a subspace of the total Hilbert space \cite{QY17}. Here the  evolution of the region supporting the operator $O_{e_1}$ is identified with the evolution of physical states, since they are on the same graph or network. In this particular situation, we thus see the important role played by non-unitary evolutions of a subspace in describing the local causal structure. Now the tensor network representation for quantum causal histories (e.g. Figure \ref{fTVD}) has the following advantage: each bond represents a non-unitary CPTP map, so that the local light-cone structure can always defined on each event in quantum causal histories.

Now that the local light cone can be defined on each node of the tensor network for the temporally varying discretizations, we can obtain a global causal future $J^+(e)=\cup_k\text{Lightcone}(e_k)$ of an event $e$  as the succession of local light cones. This $J^+(e)$  extends in principle all the way to the the future infinity $J^+$ if the network can be infinitely extended. If the succession of local light cones stops at an event $e_h$, then the events before $e_h$ will not have any influence on the network after $e_h$. In this sense, the $e_h$ bounds a region in the quantum causal histories that does not affects its causal future, and this region is a black hole region when viewed from  $J^+$. 
\begin{figure}[t]
\[
\xymatrix{
\dots\ar@{-}[r] & *+[F]{K} \ar@{-}[r]\ar@{.}[d] &  *+[F=]{P'}\ar@{-}[r] &  *+[F]{K}\ar@{-}[r]\ar@{.}[d]&\dots\\ 
\dots\ar@{-}[r] & *+[F]{K} \ar@{-}[r]_h &\ar@{-}[r] &  *+[F]{K}\ar@{-}[r]&\dots
 }
\]
\caption{The history state has unmatched constraints near the horizon event $e_h$. }\label{f99}
\end{figure}
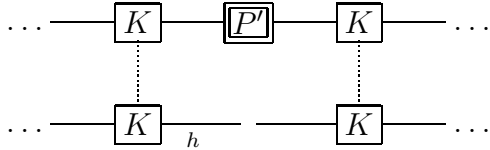
In the tensor network representation, the black hole region can be easily identified by the broken links at $e_h$. See for example Figure \ref{f99}.

This simple identification of black hole region is due to the fact that the a quantum causal history state describes the local evolution of a subsytem, instead of the global evolution of a complete Cauchy surface. The part of the Cauchy surface that falls into a black hole cannot influence the future infinity, but the entanglement between histories remains, which further implies the possibility of information transfer from a black hole.
\section{Concluding remarks}\label{sec5}
We have constructed the  tensor network representations for quantum causal histories with local non-unitary open quantum system evolutions. The main idea is using the channel-state duality of CPTP maps to transform the CPTP maps describing the local dynamical evolutions of a quantum causal history to bipartite entangled states that are suitable for tensor network constructions. We have also constructed the tensor networks for entangled quantum causal histories and holographic relations in some restricted cases. An example has been found in fully constrained quantum systems on temporally varying discretizations.  These tensor networks are constructed without any guidance of a dual QFT, thereby providing a direct bulk description of quantum gravity states.

We have spent most of the time on motivating and constructing these definitions. To apply the tensor networks defined here, one faces the problem of relevance of this construction on causets, in view of the lack of examples in most of the models of quantum gravity. Although there are strong reasons to study quantum gravity or quantum systems on discrete structures by local unitaries \cite{AM17}, the non-unitary time evolutions generically arise when we take an internal point of view that we can only probe the open quantum system with no knowledge on the ``environment" part. The decomposition of global dynamical evolutions into local unitaries is  an external approach in the same sense of the quantum simulation by quantum circuits that it requires an external agent to perform the decomposition and keep track of all possible correlations between subsystems. The local, or internal, approach to quantum gravity returns to the old philosophy of {\it relative formulation} of a quantum theory for the universe as a whole \cite{Eve57}. Here we have used the more workable theory of open quantum systems, avoiding interpretational issues.

An obvious problem in the construction of tensor networks for entangled quantum causal histories is that the causal orders are allowed to be superposed in a fully quantum theory. This quantum superposition of causality will lead to indefinite causal structure, which has been shown in recent years to be very useful in quantum information and computation. See for example \cite{Bru14}. Here we have base the theory of quantum causal histories on causets which are  defined  by definite causal structure. The apparent inconsistency between causets and quantum theory with indefinite causal structure suggests us to look for other relational frameworks of quantum gravity. But the tensor network constructions need to be changed accordingly. Another problem concerns the technical assumption of internal causality which makes the theory mathematically more manageable but hinds some structures such as the quantum common causes \cite{QCM17}. To relate the current model of spacetime to the quantum causal models studied in quantum foundation is an interesting topic for future investigations.

Based on previous works on holographic tensor networks, one can think of calculating the entanglement entropy of the constructed tensor networks. We note that the MPS of a single quantum causal history is indeed a superstate in the sense of \cite{CJQW18}. Therefore the entanglement entropies should be the spacetime entropies of superdensity operators, instead of the  R\'enyi entropy. It is still unclear how to combine the spacetime entropy and the causet structure to reach a holographic entanglement entropy relation.

The temporally extended tensor network states for quantum causal histories, in addition to the relations with superstates, also have many similarities with the precess tensor approach to open quantum systems. The process tensor formalism has been recently applied to the causal tensor networks, which can be interpreted as the path integral geometries on curved spacetime. Cf. the recent works \cite{MV18,JP19}. The process tensor formalism also has many similarities to the quantum causal models \cite{QCM17}. 
Given the usages of channel-state duality  in these models, it is promising to  relate these different formalisms.

Finally, we ask if the network structure has any guidances for constructing the quantum states (of gravity). In principle, the answer is yes, as abstractly revealed by \eqref{520} in Appendix \ref{AppB}. More practically, however, we need to work out the rules for the network structures to affect the quantum states. A plausible set of rules have been studied in \cite{BCJ11}. In this way, we hope to see how the causal structure affects the quantum states (of gravity) via tensor networks.
\section*{Acknowledgements} 
I thank the anonymous referees for their insightful comments.
 This work is partially supported by the National Natural
Science Foundation of China  through the Grant
Nos. 11875006 and  11961131013.

\appendix
\section{Quantum causal history state as  superstate}\label{AppA}
In the main text, the MPS $\ket{\mathsf{h}}$ \eqref{10} of a single quantum causal history  $\mathsf{h}$ encodes the multi-event correlations along $\mathsf{h}$. This kind of states, although entangled, are different from the entangled consistent history states that are  superpositions of histories on different spatial trajectories. But we show in the following that the MPS form of $\ket{\mathsf{h}}$ certifies them as  superstates, the generalization of the entangled history states to the superdensity operator formalism \cite{CJQW18}.

Superstates or superdensity operators generalize the entangled history states by replacing the (position) projection operators in the class operators by general operators. A simple illustration is given by a $d$-dimensional  main quantum system coupled to $n$ $d^2$-level auxiliary quantum systems. Suppose the initial state of the main system is $\ket{\psi_0}$. Consider  the unitary evolutions of the main system where the $i$-th step of the evolution corresponds to unitarily transforming the $i$-th auxiliary system, e.g. $\ket{0}_i\mapsto\sum_j\ket{j}_i$, then if the joint initial state is $\ket{0}_1\otimes...\otimes\ket{0}_n\otimes\ket{\psi_0}$, we obtain, after $n$ steps of evolution, the superstate
\begin{equation}
\ket{\Psi}=\frac{1}{d^{n/2}}\sum_{i_1i_2...i_n}\ket{i_1,i_2,...,i_n}\otimes X_{i_n}U(t_n,t_{n-1})X_{i_{n-1}}...X_{i_2}U(t_2,t_1)X_{i_1}\ket{\psi_0}
\end{equation}
where $U$ is the unitary evolution operator of the main system, and the $X_i$'s are operators on the Hilbert space $H_i$ of the $i$-th auxiliary system. 
By tracing out the state of the main system, one obtains the superdensity operator (of the auxiliary systems)
\begin{align}
 \varrho=\frac{1}{d^{n}}\sum_{i_1i_2...i_n, j_1j_2...j_n}&\text{tr}\bigl( X_{i_n}U(t_n,t_{n-1})X_{i_{n-1}}...X_{i_1}\rho_0X^\dag_{j_1}U^\dag(t_2,t_1)X^\dag_{j_2}...X^\dag_{j_n}\bigr)\cdot\nonumber\\
&\cdot\ket{i_1,i_2,...,i_n}\bra{j_1,j_2,...j_n}
\end{align}
where $\rho$ is the density matrix of the initial state of the main system. Mathematically, such superdensity operators are in fact maps $\varrho:B^*(H)\times B(H)\rightarrow\mathbb{C}$ from bounded operators on $H$ to complex numbers. Therefore, the unitary evolution operators can actually be replaced by CPTP maps between the algebras $\varphi_i:B(H_i)\rightarrow B(H_{i+1})$. In this way, we get the superdensity operator with CPTP maps,
\begin{equation}
 \varrho=\frac{1}{d^{n}}\sum_{i_1i_2...i_n, j_1j_2...j_n}\text{tr}\bigl( X_{i_n}\varphi_{n-1}[...\varphi_1[X_{i_1}\rho_0X^\dag_{j_1}]...]X^\dag_{j_n}\bigr)\ket{i_1,i_2,...,i_n}\bra{j_1,j_2,...j_n}.
\end{equation}

Let us turn to the single-history MPS $\ket{\mathsf{h}}$ \eqref{10}. 
 In the construction of \eqref{10}, we have assumed a bipartite maximally entangled states $\sum_i\ket{\psi_i}\ket{\psi_i}$  on each pair of causally related events, and the channel-state duality transforms this into $\sum_i\ket{\psi_i}\phi^\dag(\ket{\psi_i})$. At each event, we have used the projection operator 
\begin{equation}
\beta_l=\sum_{ijk}M^{(l)}_{ijk}\ket{\psi_k^{(l)}}\phi^\dag(\bra{\psi_{i}})\bra{\psi_{j}}
\end{equation}
to consistently glue the causal evolutions and give the states $\ket{\psi^{(l)}}$. Based on these features,
 we can choose the quantum causal history $\mathsf{h}$ as the main system and the states $\ket{\psi^{(l)}}$ as auxiliary. But the initial state $\ket{\psi_{\text{ref}}}$ of the system should be $n-1$ pair of maximally entangled bipartite reference states on the causal relations for neighboring events, and then the evolutions as CPTP maps transform an entangled pair via the channel-state duality
\begin{equation}
\sum_i\ket{\psi_i}\ket{\psi_i}\mapsto\sum_i\ket{\psi_i}\phi^\dag(\ket{\psi_i}).
\end{equation}
The projection operators $\beta_l$ finally act on the states of the main system to give the auxiliary states  $\ket{\psi^{(l)}}$. In this way, we obtain the single-history state $\ket{\mathsf{h}}$ as a superstate,
\begin{equation}\label{A24}
\ket{\mathsf{h}}=\frac{1}{N}\sum_{i_1i_2...i_n} \beta_n\phi^\dag_{(n-1)n}\beta_{n-1}...\beta_2\phi^\dag_{12}\beta_{1}\ket{\psi_{\text{ref}}}
\end{equation}
where $N$ is a normalization factor, and $\beta_1=\sum_{jk}\ket{\psi^{(1)}}\bra{\psi_{1,j}}, \beta_n=\sum_{jk}\ket{\psi^{(n)}}\bra{\psi_{n,j}}$. Then substituting all terms into \eqref{A24} we retain \eqref{10}.

Notice that in $\beta_l$ the tensorial coefficients $M_{ijk}^{(l)}$ can be interpreted as codifying the transformations from the evolved states $\phi^\dag(\ket{\psi_i})$ to the reference state $\ket{\psi_{j}}$ and at the same time projecting them to the auxiliary $\ket{\psi_k^{(l)}}$. This is in line with the superdensity operator formalism where an arbitrary unitary operator $U_i$ on the  Hilbert space $H_i$ of the main system  is transformed to the orthonormal basis operator $X_i$ of such operators  through a linear transform, e.g.
\begin{equation}\label{A25}
 \sum_{ij}\text{tr}(X_iU^\dag_j)U_j=\sum_iX_i.
\end{equation}
We can see that the transform coefficients in \eqref{A25} are exactly tensorial contractions. 

Now it is immediate to calculate the spacetime entropies \cite{CJQW18} of quantum causal histories. A point to keep in mind is that for quantum causal histories the spacetime geometry has not been constructed yet, so these entropies can be interpreted as pre-geometric or purely quantum-gravitational results.

Another interesting point is that the single-history superstate \eqref{A24} allows us to directly relate its causal relations to the quantum causal influence defined in \cite{CHQY18}. To see this, let us write down the formal superdensity operator of a single-history superstate,
\begin{equation}
 \varrho_{\mathsf{h}}=\frac{1}{N^2}\sum_{i_1i_2...i_n, j_1j_2...j_n}\text{tr}\bigl( \beta_{i_n}\phi^\dag_{n-1}[...\phi^\dag_1[\beta_{i_1}\rho_{\text{max}}\beta^\dag_{j_1}]...]\beta^\dag_{j_n}\bigr)\ket{\psi_{i_1},...,\psi_{i_n}}\bra{\psi_{j_1},...,\psi_{j_n}}
\end{equation}
where $\rho_{\text{max}}$ is the density matrix the maximally entangled initial reference state $\ket{\psi_{\text{ref}}}$. Since the CPTP maps only form a dynamical semigroup, these $\phi^\dag$'s cannot be canceled as for unitary evolution operators ($UU^\dag=1$) in the trace. Thus, the correlation functions involving CPTP maps are definitely related to the causal order in $\mathsf{h}$, making the quantum causal influence non-vanishing for every quantum causal history.
\section{Categorical perspectives}\label{AppB}
Category theory first comes into play when  Isham generalizes the quantum logic of consistent histories to the internal logic, i.e. topos, of consistent histories \cite{Ish97}. It is shown  that the logical algebra of propositions in consistent histories is neither a Boolean algebra nor a quantum logic, but a Heyting algebra in a suitable topos. Similar structures in causets have been studied in \cite{Mar00b}. For quantum causal histories with states living on the events, the Heyting algebra of logical propositions no longer holds, since the  superpositions of quantum states are not distributive. One might expect a more general topos for quantum causal histories.
But before that, there is a question of finding the suitable categorical description of local causal histories with non-unitary open quantum system dynamics.
Here we restrict ourselves to this question, and put aside the topos theory.

First recall that the quantum causal histories can be obtained as a functor 
\begin{equation}
Q:{\bf PSet}_{\mathcal{C}}\rightarrow{\bf Hilb}
\end{equation}
from the category of posets underlying the causet $\mathcal{C}$ to the category of finite-dimensional Hilbert spaces. The defining properties of a causet are thence required to be functorial with respect to $Q$ in the sense that they persist for Hilbert spaces on events.  This way we can alternatively work with a functor 
\begin{equation}
\tilde{Q}:{\bf CSet}_{\mathcal{C}}\rightarrow{\bf Hilb}_\mathcal{C}
\end{equation}
that preserves the causal structure, where ${\bf CSet}_\mathcal{C}$ is the category of events and causal relations in a causet $\mathcal{C}$ and ${\bf Hilb}_\mathcal{C}$ is the category of Hilbert spaces over the events of $\mathcal{C}$ and the causal evolutions in between.

We can, as in \cite{Mar00b},  consider the past $P(p)=\bigcup_i\mathsf{h}_i(p)$ of an event $p$ which is a subset of the causet $\mathcal{C}$. Suppose there exists an initial event for each history, then we can form the category ${\bf HP}_\mathcal{C}$ of the histories $\mathsf{h}_i(p)$ in the pasts $P(p)$ of events  $p\in\mathcal{C}$ and the maps between them that preserves the causal relations. These maps can chosen to be those defining a graded structure: the ``{\it face map}" that deletes an event but keeps the causal orders, which is ensured by the transitivity condition; the ``{\it degeneracy map}" that adds an event $p$ at $p$ such that $p\leqslant p$, which is just the reflexivity. Let the degree of a single quantum causal history $\mathsf{h}$ be the number $\mathsf{n}$ of events in it, then we have a simplicial set
\begin{equation}
S:{\bf HP}_\mathcal{C}^{\text{op}}\rightarrow{\bf Set}
\end{equation}
where  {\bf Set} denotes the category of sets, and the graded structure in  ${\bf HP}_\mathcal{C}$ is labeled by $\mathsf{n}$. Clearly, the category ${\bf HP}_\mathsf{h}$ of a single history $\mathsf{h}$ is equivalent to a poset category ${\bf PSet}_\mathsf{h}$, because the causal order is preserved by morphisms introduced above. Then on each single history, we also have a simplicial set 
\begin{equation}
S_\mathsf{h}:{\bf HP}_\mathsf{h}^{\text{op}}\rightarrow{\bf Set}
\end{equation}
which comes from the nerves of the posets in ${\bf PSet}_\mathsf{h}$.

The quantum counterpart of a history is of course $Q_\mathsf{h}:{\bf PSet}_\mathsf{h}\rightarrow{\bf Hilb}$. For multiple histories of $p$ in $P(p)$, each of them can be lifted to a quantum causal history by the functor $Q_\mathsf{h}$, but they can still intersect or overlap each other.
 The important point here is that for multiple quantum causal histories, the assumption of internal causality turns the causal relations between two events into a poset. Therefore, the category ${\bf Hilb}_{\mathcal{C}}$ becomes a { poset-enriched category}. Using again the nerve construction, we can turn ${\bf Hilb}_{\mathcal{C}}$ into a simplicially enriched category. ${\bf Hilb}_{\mathcal{C}}$ is furthermore a 2-category with two poset categories: one  in the direction of causal evolution, and the other in the enrichment.

A simplicially enriched category can be converted into a differential graded (dg) category by using the cobar construction for many objects \cite{Por08}. Without going into details, we can already see that a dg category is indeed a suitable way for describing the temporally evolving quantum causal histories. Intuitively, the differential in a dg category changes the degree of its objects or the number of events in a single quantum causal history, while the dg structure leads to the causet structure.

We turn to the next question of describing the open system dynamics for local quantum causal histories. To this end, one needs to select a particular subsystem and its time evolutions from many subsystems and keep the causal order at the same time, for which the lax functor is a suitable notion \cite{Dug18}.

 Consider again the 2-category ${\bf Hilb}_{\mathcal{C}}$ of quantum states and causal evolutions on  histories. We recall from Definition \ref{d21} that the CPTP maps, when viewed from the algebras on events, take the reversed direction as compared to those on states. That is, given two causally related events $p,q$ with $p\leqslant q$ in a causet $\mathcal{C}$, a CPTP map acts on the algebras on them as $\phi:\mathcal{U}(q)\rightarrow\mathcal{U}(p)$, and acts on states as  $\phi^\dag:H(p)\rightarrow H(q)$. The assumption of internal causality should also be imposed on the algebras on an coinciding event $p$ with a reversed order, e.g. $\mathcal{U}(p')\subseteq\mathcal{U}(p)$ for $p\curlyeqprec p'$. In this way, the poset enrichment can be defined by the reversed partial order on algebras. We call the reversed order oplax, so that the following constructions on ${\bf Hilb}_{\mathcal{C}}$  are lax.

We can define the category {\bf D} of ``dynamics" by lax functors and lax natural transformations as follows.  The objects of {\bf D} are the lax functors
\begin{equation}
L:{\bf HP}_\mathcal{C}\rightarrow {\bf Hilb}_{\mathcal{C}},
\end{equation}
and the morphisms are the lax natural transformations between the lax functors, which we denote by $\dot{L}:L\looparrowright L'$.  Similarly, on a single history $\mathsf{h}$, we have  $L_\mathsf{h}:{\bf HP}_\mathsf{h}\rightarrow {\bf Hilb}_\mathsf{h}$ and $\dot{L}(\mathsf{h},\mathsf{h}')$. Now the question is how to select from $L$ a subsystem dynamics localized on the history $\mathsf{h}$. A possible answer is given by the following lax natural transformation
\begin{equation}\label{520}
\Bigl(L:{\bf HP}_\mathcal{C}\rightarrow {\bf Hilb}_{\mathcal{C}}\Bigr)\looparrowright\Bigl(S_\mathsf{h}:{\bf HP}_\mathsf{h}^{\text{op}}\rightarrow{\bf Set}\Bigr)
\end{equation}
which picks out the subsystem dynamics according to the simplicial set on a single history $\mathsf{h}$. Note that such an abstract selection of subsystem can only reveal its classical network structure, and the details of the quantum dynamics still need further physical inputs.
\bibliographystyle{amsalpha}

\end{document}